\documentstyle[11pt,aaspp4]{article} 

\newcommand{\V}{\mbox{\it V}}
\newcommand{\J}{\mbox{\it J}}
\renewcommand{\H}{\mbox{\it H}}
\newcommand{\K}{\mbox{\it K}}

\newcommand{\VK}{\mbox{$\V\!-\!\K$}}

\newcommand{\JH}{\mbox{$\J\!-\!\H$}}
\newcommand{\HK}{\mbox{$\H\!-\!\K$}}
\newcommand{\AV}{\mbox{$A_{\it V}$}}

\newcommand{\AK}{\mbox{$A_{\it K}$}}

\newcommand{\MH}{\mbox{$M_{\it H}$}}

\newcommand{\mv}{\mbox{$m_{\it V}$}}
\newcommand{\mk}{\mbox{$m_{\it K}$}}
\newcommand{\mbol}{\mbox{$m_{\rm bol}$}}

\newcommand{\Msun}{\mbox{$M_\odot$}}
\newcommand{\Lbol}{\mbox{$L_{\rm bol}$}}

\newcommand{\BC}{\mbox{$\rm BC$}}
\newcommand{\Teff}{\mbox{$T_{\rm eff}$}}
\newcommand{\degree}{\mbox{$^\circ$}}

\newcommand{\Te}{\mbox{$T_{\rm e}$}}
\newcommand{\Tb}{\mbox{$T_{\rm b}$}}

\newcommand{\EM}{\mbox{$E$}}

\renewcommand{\[}{\begin{eqnarray}}
\renewcommand{\]}{\end{eqnarray}}
 
\newcommand{\Br}{{\rm Br\ifmmode \gamma \else $\gamma$\fi}}

\newcommand{\g}{G29.96$-$0.02}

\newcommand{\HII}{{H\mbox{$\,$}II}}

\begin{document}

\slugcomment{Accepted to appear in ApJ}

\title{
Direct Observations of the Ionizing Star
in the UC HII Region {\g}:
A Strong Constraint on the Stellar Birth Line for Massive Stars
}

\author{Alan M. Watson}
\affil{Lowell Observatory,
1400 West Mars Hill Road,
Flagstaff, AZ 86001\\
and\\
Department of Astronomy,
New Mexico State University,
Las Cruces,
NM 88001\\
alan@oldp.nmsu.edu
}

\author{Alison L. Coil}
\affil{Lowell Observatory,
1400 West Mars Hill Road,
Flagstaff, AZ 86001\\
and\\
Department of Astrophysical Science,
Princeton University,
Peyton Hall, Princeton, NJ 08544\\
alcoil@astro.princeton.edu
}

\author{Debra S. Shepherd}
\affil{Department of Astronomy,
University of Wisconsin -- Madison,
475 North Charter Street,
Madison,
WI 53706\\
and\\
Radio Astronomy,
California Institute of Technology,
Pasadena, CA 91125\\
dss@milli.caltech.edu
}

\author{Peter Hofner}
\affil{Universit\"at zu K\"oln, I. Physikalisches Institut,
Z\"ulpicherstrasse 77, D-50937 K\"oln, Germany\\
and\\
NAIC, Arecibo Observatory,
PO Box 995, Arecibo, PR 00613\\
hofner@naic.edu}

\and

\author{Ed Churchwell}
\affil{Department of Astronomy,
University of Wisconsin -- Madison,
475 North Charter Street,
Madison,
WI 53706\\
churchwell@astro.wisc.edu}

\pagebreak

\begin{abstract}
\rightskip=\leftskip
We have observed the ultracompact HII region {\g} in the near infrared
{\J}, {\H}, and {\K} bands and in the {\Br} line. By comparison with
radio observations, we determine that the extinction to the nebula is
$\AK = 2.14$ with a 3$\sigma$ uncertainty of 0.25. We identify the
ionizing star and determine its intrinsic {\K} magnitude. The star
does not have an infrared excess and so appears to be no longer
accreting. The {\K} magnitude and the bolometric luminosity allow us
to place limits on the location of the ionizing star in the HR
diagram. The 3$\sigma$ upper limit on the effective temperature of the
ionizing star is $42\,500~\rm K$. We favor a luminosity appropriate
for star with a mass in excess of about $60~\Msun$. The limit on the
temperature and luminosity exclude stars on the ZAMS and stars within
$10^6\rm~yr$ of the ZAMS. Since the age of the UC~HII region is
estimated to be only about $10^5\rm~yr$, we suggest that this is
direct evidence that the stellar birth line for massive stars at twice
solar metallicity must be significantly redder than the ZAMS.
\end{abstract}

\section{Introduction}

Our understanding of the formation of massive stars is primitive. In
this paper, we show how observations of massive stars
in ultracompact (UC) HII regions can provide information about
their birth. UC HII regions are found around young massive stars still
embedded in their natal molecular clouds. Statistically, the objects
classified as UC~HII regions by Wood \& Churchwell (1989a, 1989b)
appear to last for about 10--20\% of the life of an O star or a
few~$\times 10^5\rm~yr$. Massive stellar outflows (Harvey \& Forveille
1988; Garden \& Carlstrom 1992; Hunter et~al.\ 1995; Shepherd \&
Churchwell 1996a, 1996b) seem to represent a still earlier and more
poorly understood phase, possibly during which the massive star is
still accreting.

Previously, the ionizing stars in UC~HII regions have been studied
through their indirect effects on the surrounding gas and dust. These
studies include determinations of the bolometric luminosities (Chini
et al.\ 1986; Wood \& Churchwell 1989a), effective ionizing fluxes
(Wood \& Churchwell 1989a; Kurtz, Churchwell, \& Wood 1994), and
effective spectral hardnesses (Doherty et al.\ 1994). All of these
methods have significant drawbacks. The bolometric luminosities are
upper limits because they are determined largely from IRAS
observations and most likely include contributions from luminous
sources other than the ionizing star of the UC~HII region. The
effective ionizing flux and the effective spectral hardness are very
difficult to relate to the star because of the effects of dust within
the UC~HII region and large uncertainties in stellar ionizing continua
(see the discussion in Najarro et~al.\ 1996).

Instead, we have chosen to observe the UC~HII region {\g} in the
near infrared {\J}, {\H}, and {\K} bands and in the 2.17{\micron}
{\Br} line. These observations allow direct measurements of the
properties of the ionizing star. {\g} has been previously imaged in
the near infrared by Van~Buren (1993), Megeath (1993), and Fey et~al.\
(1995). These authors extracted a great deal of qualitative
information from their observations and were able to make a number of
suggestions. Our work builds on their work; we are able to confirm a
number of their suggestions and derive important new information on
the ionizing star and the nebula.

Fey et al.\ (1995) considered a champagne flow model for {\g} and
estimated an age for the UC~HII region of about $10^5~\rm yr$ from the
extent of the extended radio continuum emission. Mac Low et~al. (1991)
and Van~Buren \& Mac~Low (1992) modeled {\g} as a bow shock around a
star moving at $5$--$20~\rm km\,s^{-1}$ at an angle of about
135{\degree} to our line of sight. However, the O star is closely
associated with a young cluster (Fey et al.\ 1995 and below) and a hot
core apparently containing a protostar (Cesaroni et~al.\ 1994); this
suggests that it could have moved at most 10~arcsec and so, again, has
an age of less than $10^5\rm~yr$. Lumsden \& Hoare (1996) found that a
bow shock is a poor fit to their {\Br} velocity field and prefer
instead a model similar to the one suggested by Fey et al.\ (1995).
Mass-loaded models (Hollenbach et~al.\ 1994; Dyson 1994; Lizano \&
Cant\'o 1995; Dyson, Williams, \& Redman 1995; Redman, Williams, \&
Dyson 1996; Williams, Dyson, \& Redman 1996; Lizano et~al.\ 1996) can
delay the expansion of an UC~HII region by about $10^5~\rm yr$. If the
central star was in an outflow/accretion phase before the UC~HII
region was created, this phase is likely the have lasted less than
$10^4~\rm yr$, based on comparison with other O stars currently
undergoing or just finishing molecular outflow activity (Cabrit \&
Bertout 1992; Shepherd \& Churchwell 1996b; Acord, Walmsley, \& Churchwell 1997).
All of these estimates and considerations point to an age of order
$10^5~\rm yr$ for the UC HII region.
 
We adopt a heliocentric velocity of $95~\rm km\,s^{-1}$, the average
global radio recombination line velocity (Afflerbach et al.\ 1994).
Since {\g} lies close to a tangent point, its distance is uncertain,
and, unfortunately, the formaldehyde absorption line data of Downes et
al.\ (1990) do not resolve the distance ambiguity. The Galactic
rotation curve and error bars given by Burton (1988) yield a distance
of between 5 kpc and 10 kpc.

Afflerbach, Churchwell, \& Werner (1996) have determined that the
abundances of O, N, and S in {\g} are about twice those observed in
the local ISM. This is consistent with its location towards the center
of the galaxy.

The organization of this paper is as follows. In section 2 we present
our new infrared observations. In section 3 we summarize the existing
radio observations that have been kindly made available to us by a
number of our colleagues. In section 4 we present the astrometric
calibration of the location of the ionizing star. In section 5 we
discuss the bolometric luminosity of {\g}. In section 6 we investigate
the nebula using the {\Br} and radio images and derive the extinction
to the nebula. In section 7 we investigate the properties of the
ionizing star and its associated stellar cluster using the broad band
{\J}, {\H}, and {\K} images. In section 8 we summarize the principal
results of this work and discuss some of their implications.

\section{Infrared Data}

\subsection{Observations}

We observed {\g} on the night of 1995 May 21 and June 24 UT with the
OSIRIS near infrared camera (Depoy et~al.\ 1993) on the Perkins 1.8
meter telescope at Lowell Observatory. Light cirrus was present during
the first night. The second night was photometric. The seeing was 1.8
arcsec FWHM on the first night and 2.0 arcsec FWHM on the second
night.

OSIRIS has a 256$\times$256 NICMOS--3 detector. The re-imaging optics
gave a pixel scale of 0.62 arcsec and a field of view of 150 arcsec.
Images were taken through {\J}, {\H}, and {\K} broad band filters and
2.12, 2.14, and 2.17 {\micron} narrow band filters. The 2.12 {\micron}
filter includes emission from the H$_2$ $\nu=0$--$1$ S(1) line and the
2.17 {\micron} filter includes emission from the {\Br} line; the 2.14
{\micron} filter excludes strong emission lines.

We took several exposures through each filter, dithering the target
onto each of the detector quadrants. Total exposure times in each of
the broad band filters were 120 seconds on the first night and 240
seconds on the second night. Total exposure times in each of the
narrow band filters were 480 seconds on the first night and 240
seconds on the second night.

\subsection{Reduction}

The data were reduced using custom programs within the VISTA image
reduction system. Each image was corrected for non-linear response by
subtracting a bias image and dividing each pixel by its estimated
relative response. We have found that the non-linearity can be
adequately modeled as a quadratic function and the principal variation
in non-linearity in OSIRIS is between quadrants, so we determined and
applied different quadratic coefficients for each quadrant. Each image
was then dark subtracted, flattened, and masked to eliminate bad
pixels. A single sky value was estimated for each image by fitting a
quadratic to the peak of the intensity histogram and solving for the
turning point. That value was then subtracted from the image. The
images were brought into a common registration by measurements of
bright stars, shifted by an integral number of pixels, and co-added.
This process did not noticeably degrade the seeing.

We corrected the 2.12 and 2.17 {\micron} images for continuum emission
by subtracting the 2.14 {\micron} image after scaling by the ratio of
the mean brightnesses of several stars away from the UC {\HII} region.
We did not detect any line emission in the 2.12~{\micron} image and
will not consider it further.

\subsection{Broad Band Calibration}

The broad band data from the second nights data were calibrated by
observations of the two UKIRT faint standards 25 and 28 (Casali \&
Hawarden 1992). The UKIRT faint standards contain very little color
information, so we did not attempt to solve a color term and have left
the magnitudes on the natural system. This is less of a drawback than
it seems, as our sources are so heavily reddened that color terms
derived from observations of unreddened stars would be of limited use.
The RMS residuals in the fits for the zero points and extinction terms
were about 3\%.

\subsection{Narrow Band Calibration}

The narrow band data from the second night was calibrated by
observations of the sdO optical spectrophotometric standard
BD~+28~4211. This star is not an infrared spectrophotometric standard
but we expect that its spectrum will be featureless in the {\K}
window. BD~+28~4211 has a fainter red companion 3 arcsec to the NNW
(Massey \& Gronwall 1990; Thejll, Theissen, \& Jimenez 1994). We
measured a difference between the fainter and brighter stars of
$\Delta K \approx 1.5$. Before performing photometry, we
removed the companion by shifting, scaling, and subtracting a suitably
centered star. We measured a {\K} magnitude of $11.56 \pm 0.04$ for
BD~+28~4211. The absolute calibration of Bessell \& Brett (1988) gives
a flux of $9.62 \times 10^{-16}~\rm erg\,s^{-1}\,cm^{-2}\,\AA^{-1}$ at
2.19~{\micron} with a $1\sigma$ uncertainty of about 4\%. Assuming a
$\lambda^{-4}$ spectrum, this gives a flux of $1.01 \times
10^{-15}~\rm erg\,s^{-1}\,cm^{-2}\,\AA^{-1}$ at 2.164~{\micron}, the
peak of the 2.17~{micron} filter.

\subsection{Infrared Images}

Figure~\ref{fig-ir} presents images in {\J}, {\H},
{\K}, and {\Br}. The arc-shaped nebula at the center of each image is
the UC~HII region {\g}. The bright point source within the nebula in
the {\J}, {\H}, and {\K} images is the ionizing star candidate
previously seen by Van~Buren (1993), Megeath (1993), and Fey et~al.\
(1995). The position of this source is marked with a white cross in the
{\Br} image.

\section{Radio Data}

{\g} has been extensively studied in the radio continuum, radio
recombination lines, and molecular lines. We are grateful to a number
of our colleagues for providing the radio images listed in
Table~\ref{tab-radio-data} in electronic form.
Table~\ref{tab-radio-data} lists for each image the authors,
frequency, telescope, beam FWHM, and largest angular size imaged. The
VLA images are all of radio continuum emission. The OVRO image is the total
intensity in the C$^{18}$O $J=1-0$ line. We assume that the 3$\sigma$
errors in the absolute flux calibration of the radio continuum images
are about 20\%. The errors in the C$^{18}$O image are larger but do
not affect our analysis.

\section {Astrometry}

We determined the position of the ionizing star candidate (\#3 below) by
cross-correlating versions of the {\Br} image with the
astrometrically-calibrated 1.3 and 2 cm radio continuum images. Before
cross-correlating, we smoothed each pair of images to the same
resolution. The ionizing star candidate is located at $\alpha_{1950} =
18~43~27.214$ and $\delta_{1950} = -02~42~35.9$ in the radio reference
frame. The determinations from the 1.3~cm and 2~cm images differed by
only 0.2 arcsec. The distance between the ionizing star and the peak
of the 2~cm emission is 2.0 arcsec. At 5.0~kpc this corresponds to
10000~AU.

\section{The Bolometric Luminosity}
\label{sec-lbol}

Table~\ref{tab-sed} lists measurements of the continuum spectral
energy of {\g}. The values at 21~cm, 6~cm, and 1.3~cm were measured by
us from the VLA maps obtained by Claussen \& Hofner (1996), Afflerbach
et al.\ (1994), and Cesaroni et al.\ (1994). The three measurements at
1.3~mm were obtained with different effective beam sizes. The
measurements at 790~{\micron} and 350~{\micron} from Hunter (1997) may
be missing extended emission. To obtain total magnitudes of $\J =
12.5$, $\H = 10.1$, and $\K = 7.9$ for the nebula we measured the
total magnitudes within a 25~arcsec box centered on the ionizing star
after subtracting scaled point spread functions of all stars except
the ionizing star. (We describe our photometry below.) We converted
these to Jy using the flux calibration of Bessell \& Brett (1988). Our
measurements of $\Delta J \approx 1.9$, $\Delta H \approx 1.8$, and
$\Delta K \approx 2.4$ between the flux for the ionizing star and the
total flux show that nebular emission and scattering dominate over
direct light from the star even at these short wavelengths.

Figure~\ref{fig-sed} shows the spectral energy distribution plotted as
$\log \nu F_\nu$ against $\log \nu$. The spectral energy distribution
is characteristic of UC HII regions: optically thick free-free
emission longward of a few cm, optically thin free-free emission
between a few cm to about a few mm, and very strong thermal dust
emission from about 1~mm to about 10~{\micron}.

We can estimate the bolometric luminosity by integrating over the
spectral energy distribution. As expected, extrapolation of the cm
wavelength free-free emission to 1300~{\micron} (the dashed line in
Figure~\ref{fig-sed}) indicates that the 1300~{\micron} emission
arises from hot dust. We fitted a modified Planck law $F_\nu \propto
\nu^2 B_\nu(T)$ to the 1300~{\micron} and 100~{\micron} points,
finding $T \approx 26~\rm K$, and interpolated elsewhere. This adopted
spectrum is shown by a solid line in Figure~\ref{fig-sed}. We also
considered the optically thick model fit by Hunter (1997); this gives
a flux within 5\% of ours. We derive an apparent bolometric flux of
$7.00 \times 10^{-7}~\rm erg\,s^{-1}\,cm^{-2}$ which corresponds to an
apparent bolometric magnitude of $m_{\rm bol} = 3.90$ (where $M_{\rm
bol,\odot} = 4.75$). Since this depends heavily on the IRAS
100~{\micron} measurement, we assume a 3$\sigma$ uncertainty of 30\%.

Since this bolometric luminosity is integrated over an arcmin sized
region, it is likely to include contributions from other sources and
so is a priori only an upper limit for the bolometric luminosity of
the ionizing star. However, there are no other bright sources in
either the 1300~{\micron} map of Mooney et~al.\ (1995) or the mid
infrared images of Ball et~al.\ (1996). We do know that the IRAS beam
includes both the hot core to the west of the UC HII region and a
cluster of B stars apparently associated with the ionizing star (see
below). Cesaroni et~al.\ (1994) estimate that the source powering the
core contributes only about 1/10 of the flux. Hofner et al.\ (1997)
estimate that at 2.7~mm only $150 \pm 50~\rm mJy$ of the total
$2.73~\rm Jy$ arises from the hot core. Much of the total emission
will be free-free emission from the UC~HII region, but since the hot
core does not appear in cm wavelength continuum maps we attribute its
emission solely to dust. We estimate the total dust emission of the
region at 2.7~mm to be about 4~Jy by extrapolating the 1.3~mm
measurement of 15.6~Jy (Mooney et al.\ 1995) assuming a $\nu^2$
spectrum. Thus, the hot core contributes less than 10\% of the 2.7~mm
dust emission. We expect that the hot core has a spectrum that is no
hotter than the UC~HII region and so this fraction is an upper limit
to its total contribution to the bolometric luminosity. Furthermore,
as we shall see below, the cluster appears to be very much less
luminous than the ionizing star. This suggests that the true
bolometric luminosity of the ionizing star is likely to be quite close
to the total bolometric luminosity of the region. We will
conservatively assume that at least half of the measured luminosity
must come from the ionizing star, that is $3.90 \le \mbol \le 4.65$
with a $3\sigma$ uncertainty on the limits of 0.3.

\section{The Nebula}

\subsection{Emission Measure and Electron Temperature}

In order to predict the intrinsic flux in {\Br}, we need to know both
the emission measure and the electron temperature in the nebula. We
can use the 2~cm, 6~cm, and 21~cm radio continuum images to provide
this information. The familiar expression for the brightness
temperature {\Tb} at a given frequency $\nu$ is \[\Tb = \Te(1 -
e^{-\tau}).\] The optical depth is given by \[\tau = 8.235 \times
10^{-2} a \Te^{-1.35} (\nu/{\rm GHz})^{-2.1} (\EM/{\rm cm^{-6}\,pc})\]
where $E$ is the emission measure and $a$ is a weak function of $\nu$
and {\Te} (Mezger \& Henderson 1967). The peak synthesized beam brightness
temperature measured in the 2~cm image is about 1500~K. Anticipating
our result that {\Te} is about 6000~K and assuming that there is no
unresolved structure in the 2~cm image, equation~1 implies that the
peak optical depths at 2~cm, 6~cm, and 21~cm are about 0.3, 3, and 40.
Thus, the 6~cm and 21~cm images provide information on different
regions of the nebula. Since the dependence of $a$ on {\Te} is so
weak, we again anticipate our result and use its value at 6000~K.
 
We smoothed the 2~cm image to match the resolution of the 6~cm and
21~cm images and then calculated {\Te} and {\EM} iteratively. We
started with an initial guess for {\Te} of 6000~K. We then used the
2~cm image and equations~1 and 2 to predict the 6~cm or 21~cm image.
Half of the ratio between the predicted and actual 6~cm or 21~cm image
was used at each point in the image as a correction factor for the
value of {\Te}. We repeated the prediction/correction step until {\Te}
no longer changed significantly. Images of {\Tb} and {\Te} are shown
in Figure~\ref{fig-te}. The effects of increasing optical depth at
longer wavelength are apparent as increases in {\Tb} and a severe
reduction in contrast in the 21~cm {\Tb} image compared to the 2~cm
{\Tb} image. In the optically thick part of the nebula, {\Te} is
determined largely by the brightness temperature in the longer
wavelength image and so has a systematic 3$\sigma$ error of about
20\%. In the optically thin parts of the nebula, it is difficult to
separate the product of {\Te} and {\EM} and the errors in {\Te} are
much larger. Random errors from imperfect imaging are more difficult
to judge. The hot regions along the NW and SW of the arc in
Figure~\ref{fig-te}c and the `bumps' in Figure~\ref{fig-te}f are
almost certainly artifacts. Ignoring these, {\Te} is between 5000~K
and 7000~K over the bright part of the nebula, including the arc, and
might possibly be about 1000~K cooler in the tail to the NE. In our
following analysis we will adopt a uniform temperature of $6000~K$ and
consider temperature variations in our discussion.

Afflerbach et al.\ (1994) examined the nebular properties by a non-LTE
analysis of radio recombination lines. They derived {\Te} in the range
6200--8600~K, although with large errors. Thus, our measurement of
{\Te} of about 6000~K is in rough agreement.

\subsection{The Extinction To The Nebula}

Figure~\ref{fig-a}a shows our {\Br} image of {\g}. The total {\Br}
flux in Figure~\ref{fig-a} is $8.42\times10^{-12}~\rm
erg\,s^{-1}\,cm^{-2}$ with a $3\sigma$ uncertainty of about 12\%. We
measure $2.23 \times 10 ^{-12}~\rm erg\,s^{-1}\,cm^{-2}$ in a 5 arcsec
aperture centered on the peak of the {\Br} emission, consistent with the flux
of $2.49 \times 10 ^{-12}~\rm erg\,s^{-1}\,cm^{-2}$ measured by
Doherty et al.\ (1994), and $6.6 \times 10^{-12}~\rm
erg\,s^{-1}\,cm^{-2}$ in a 17 arcsec aperture, consistent with the flux of
$7.6 \pm 1.0 \times 10^{-12}~\rm erg\,s^{-1}\,cm^{-2}$ measured by
Herter et~al.\ (1981).

Figure~\ref{fig-a}b shows the 2~cm image of Fey et al.\ (1995) at its
full resolution and Figure~\ref{fig-a}c shows it smoothed to the
1.8~arcsec FWHM resolution of our {\Br} image. As we described above,
the alignment of the 2~cm image was determined by cross-correlation
with the {\Br} image, after smoothing the 2~cm image to match the
resolution of the {\Br} image. The total 2~cm flux density in
Figure~\ref{fig-a} is 4.6~Jy. Fey et al.\ (1995) quote a flux of
3.9~Jy for the whole image, but this includes large regions with
small, negative flux. The similarities between the {\Br} and 2~cm
images are striking; both show the strong arc of emission that lead
Wood \& Churchwell (1989a) to classify {\g} as a cometary UC HII
region and both show emission behind the arc and in extensions to the
south east and the north. The importance of this is that it
demonstrates that the emission occurs at similar optical depths and so
the extended emission noted by Fey et al.\ (1995) is directly
associated with the UC~HII region rather than being a chance
superposition of foreground or background emission.

We now derive a relation between the 2~cm brightness temperature and
the intrinsic flux in {\Br}, with the purpose of comparing the
intrinsic {\Br} flux to the observed {\Br} flux to derive the
apparent extinction at {\Br}. Unless otherwise noted, all units are cgs.
From equations~1 and 2, the
emission measure {\EM} in $\rm cm^{-5}$ is \[\EM = 4.72 a_{2\rm cm}^{-1}
\Te^{1.35} \nu_{2\rm cm}^{2.1} \ln\left({\Te \over \Te -
\Tb}\right).\] The {\Br} flux $S_{\Br}$ from a region of solid angle
$\Omega$ is \[S_{\Br} = 0.9 h\nu_{\Br} \alpha^{\rm eff}_{\Br} {\Omega
\over 4\pi}\EM.\] The factor of 0.9 arises because we have assumed
that the nebula consists of 10\% singly ionized helium by number; the
ionized helium will contribute electrons and ions to the radio
continuum emission but only electrons to the {\Br} emission. Over the
range of temperatures of interest here, Hummer \& Storey (1987) give
\[\alpha^{\rm eff}_{\Br} = 6.48 \times 10^{-11} \Te^{-1.06}.\] We
ignore variations of $\alpha^{\rm eff}_{\Br}$ with density as they
amount to less than 1\% over the range of interest here. Combining
equations~3, 4, and 5 and replacing $a_{\rm 2cm}$ and $\nu_{\rm 2cm}$
gives \[S_{\Br} = 0.612 {\Omega \over 4\pi} \Te^{0.29}
\ln\left({\Te \over \Te - \Tb}\right).\] To understand the errors it
is worth making the approximation that $\Tb \ll \Te$. The peak
synthesized beam brightness temperature in the 2~cm image is only 1500~K, so this is a good approximation to first order. We then have \[S_{\Br} \approx 0.612 {\Omega \over 4\pi}
\Te^{-0.71} \Tb.\] It can be seen that the 20\% 3$\sigma$ systematic
uncertainty in $\Tb$ and $\Te$ and the 12\% 3$\sigma$ uncertainty in
the observed flux in {\Br} translate to a 27\% 3$\sigma$ systematic
uncertainty in the extinction.

We used equation~6 to estimate the intrinsic flux of {\Br} at each
point from the 2~cm image. The total intrinsic flux in
Figure~\ref{fig-a} is expected to be $6.72 \times 10^{-11}~\rm
erg\,s^{-1}\,cm^{-2}$ and so the apparent extinction at {\Br} is
$2.26$ mag. This extinction is an average over the whole nebula and we
quote it only for illustrative purposes; we compute images of apparent
extinction below.

The apparent extinction to the nebula is not uniform.
Figure~\ref{fig-a}d shows an image of the apparent extinction. The
apparent extinction is higher in the arc and towards the end of the
extension to the south east, reaching about 2.6 mag. In the
region behind the arc, in the vicinity of the ionizing star, the
apparent extinction drops to about 2.20 mag, although there is some
noise from continuum subtraction. The peak-to-valley contrast in the
main part of the nebula is about 0.6 mag or a factor of about 1.7. It
seems significant that the two regions of highest apparent extinction
correspond to the two brightest regions. We can think of six possible
explanations for the correlation of higher apparent extinction with
surface brightness: an error in the 2~cm map, the presense of dense
ionized regions in the nebula, variations in electron temperature
within the nebula, variations in external extinction, a molecular
sheath, and internal extinction. We consider these in turn.

{(i) Errors in the 2~cm image.} The kind of error in the 2~cm image
that would explain the structure would have the characteristic of
placing almost twice as much flux in small scale structures (the arc)
than large scale structures (the tail). Since the 2~cm image is
constructed from images in three VLA configurations, such an error
might have occurred if there were relative flux errors between the
observations. To investigate this, we compared the 2~cm image to the
1.3~cm image. The peak optical depth in the 2~cm image is 0.3 and so
optical depth effects will be important at the 10\% level in the arc.
Correcting both images for optical depth effects is complicated, since
they have different intrinsic resolutions. Therefore, we compare the
real 1.3~cm image to the prediction from the 2~cm image, obtained by
assuming $\Te = 6000~\rm K$, using equations~1 and 2 to account for
optical depth and frequency effects, and smoothing from the resolution
of the 2~cm image to that of the 1.3~cm image. Figure~\ref{fig-comp}a and b
show the observed and predicted 1.3~cm images. The predicted image has
only 87\% of the flux in the observed image, but this is not
unexpected. Although there are significant variations in the ratio,
they are only at the 10\% level and are not in the sense required to
explain the additional apparent extinction (the arc does not appear
brighter in the 2~cm image than the 1.3~cm image). It would appear,
then, that errors in the 2~cm image cannot explain the variations in
apparent extinction.

{(ii) Dense Ionized Regions.}
Our procedure above accounts for optical depth effects as long as
there is no unresolved structure. It is possible, though, that
unresolved dense ionized regions exist in the nebula. These would
contribute strongly to {\Br} but might well be optically thick at
2~cm and contribute only weakly. However, in order to mimic the
variations in the apparent extinction image, these clumps would have
to avoid the brightest regions in the nebula. This seems somewhat
unlikely.

{(iii) Variations in Electron Temperature.}
As equation~7 shows, the {\Br} flux depends on the electron
temperature roughly as $T_{\rm e}^{-0.71}$. Thus, we could explain the
factor of 1.7 peak-to-valley contrast in the ratio by a factor of 2
peak-to-valley contrast in the temperature, in the sense that the
brighter regions are cooler. However, in our investigation of {\Te}
above we discovered that this was not the case: the region around the
arc is certainly not cooler than the tail and is perhaps 1000~K or
15\% hotter.

{(iv) Variations in External Extinction.}
At first glance, variable external extinction to the nebula seems a
likely candidate. Figure~\ref{fig-co} shows that {\g} is superposed
on a molecular clump with a dense core in front of the arc. However,
the contours of molecular gas do not match the details of the
variation in extinction: they would suggest a smooth gradient across
the nebula rather than the strong correlation with nebular emission
that we observe.

{(v) A Molecular Sheath.}
The bow shock model predicts a compressed molecular sheath around the
arc. Van~Buren and Mac~Low (1992) used the radio
recombination line observation of Wood \& Churchwell (1991) to show
that the velocity field implied a viewing angle of about 135{\degree}
in the bow shock model, i.e., the head is more distance than the tail.
Nevertheless, it is possible that we are viewing the edge-brightened
arc at an oblique angle through such a sheath.

We can estimate the column density through such a sheath by adopting
$A_V/N_{\rm H} = 1.0 \times 10^{-21}~\rm mag\,cm^{-2}$ and
$A_{\Br}/A_V = 0.125$ (for the $R_V = 5.0$ extinction law of
Cardelli, Clayton, \& Mathis 1989). Our value of $A_V/N_{\rm H}$ is
twice the local value to account for the higher metallicity in {\g}.
We take the local average from Bohlin, Savage, \& Drake (1978),
although Cardelli, Clayton, \& Mathis (1989) give reasons for
believing that the value may be lower in molecular clouds. The excess
$A_{\Br}$ of about 0.4~mag corresponds to $N_{\rm H_2}$ of about $1.6
\times 10^{21}~\rm cm^{-2}$. This column is similar to the predictions
of $N_{\rm H_2} \approx 5 \times 10^{21}~\rm cm^{-2}$ for the column
density through the shell in the vicinity of the stagnation point (Van
Buren et~al.\ 1990; Van Buren \& Mac Low 1992), although the
comparison is simplistic because we ignore the effects of geometry.

{(vi) Internal Extinction.}
Internal extinction would provide a natural explanation of the
correlation between excess apparent extinction and nebular emission.
We can eliminate radiation transfer effects in hydrogen by noting that
the peak emission measure in the 2~cm image is $1.4 \times 10^8~\rm
cm^{-6}\,pc$ and thus the line center optical depth in {\Br} is less
than $3 \times 10^{-7}$ (Hummer \& Storey 1987). However, dust may
provide significant opacity. We can investigate this by assuming that
dust is evenly mixed with a uniform density ionized gas in the nebula.
We ignore the effects of scattering, which may be a better
approximation than might be supposed, since the arc is most likely to
be a sheet in which scattering will effectively remove photons from
the line of sight. In this model, the internal extinction is
\[A'_{\Br} = 0.5 \left(A_{\Br}/A_V\right)\left(A_V/N_{\rm H}\right)
N_{\rm H}\]The factor of 0.5 arises because the effective optical
depth of an evenly mixed slab of emitting matter is $-\ln \int_0^1
e^{-\tau l}\,dl = -\ln\left((1 - e^{-\tau})/\tau\right) = \tau/2 +
O(\tau^2)$. If the nebula is uniform, then $N_{\rm H} = EM/n_{\rm e}$
and we can derive $N_{\rm H}$ from the 2~cm image and the electron
density of $6 \times 10^4~\rm cm^{-3}$ (the average of the values from
regions A, B, and W from Afflerbach et al.\ 1994). The peak emission
measure in the full-resolution 2~cm image is $1.4 \times 10^8~\rm
cm^{-6}\,pc$ and so this model predicts a peak effective internal
extinction of 0.41 magnitudes. However, the extinction must be
weighted by the emission and smoothed before it can be directly
compared to the {\Br} map. When this is done, the peak effective
internal extinction drops to 0.27. The emission-weighted value of
$A'_{\Br}$ is shown in Figure~\ref{fig-aint}a. Figure~\ref{fig-aint}b
shows $A_{\Br} - A'_{\Br}$, the apparent extinction after
subtracting the model internal extinction. The pattern of higher
apparent extinction is markedly reduced in the vicinity of the arc but
is not eliminated. Figure~\ref{fig-aint}c shows $A_{\Br} -
2A'_{\Br}$, the apparent extinction after subtracting twice the model
internal extinction. This almost completely removes the traces of
non-uniform extinction over the majority of the nebula. A number of
things could act to raise the internal extinction above our model. We
could have underestimated $A_V/N_{\rm H}$ by a factor of two, but this
seems unlikely (Cardelli, Clayton, \& Mathis 1989). It seems more
likely that the nebula is is not uniform in density but includes
relatively dense ionized regions with $n_e \sim 10^5 \rm cm^{-3}$ that
contribute to both the emission and extinction along with either
lower density ionized regions or neutral regions that only contribute
to the extinction.

Ultimately, our lack of an alternative explanation and the striking
correlation between the variations in extinction and the brightness of
the nebula leads us to the conclusion that we must be seeing the
effects of either a molecular sheath around a bow shock or of internal
extinction. That our model for internal extinction fails to predict
the magnitude of the extinction by a factor of two suggests that there
are lower density or neutral inclusions in the arc of the nebula with
about as much column density as the dense ionized gas. This might
naturally occur in models in which UC~HII regions ablate dense
inclusions (Dyson 1994; Lizano \& Cant\'o 1995; Dyson, Williams, \&
Redman 1995; Redman, Williams, \& Dyson 1996; Williams, Dyson, \&
Redman 1996; Lizano et~al.\ 1996) or in which the arc arises from a
photoevaporation front being driven into an inhomogeneous medium.

The location of the star slightly off the arc means that, regardless of
the origin of the additional apparent extinction associated with the
arc, the extinction to the ionizing star at {\Br} is 2.20~mag with a
3$\sigma$ uncertainty of 0.25~mag.

\section{Stellar Properties}

\subsection{Near Infrared Photometry}

Stellar photometry in the region of the UC HII region is hindered
by the strong and variable nebular background, the degree of crowding,
and the extreme colors of some of the stars which cause them to be
undetected in one or more of the three bands. We adopted methods
to minimize these problems.

First, we performed stellar photometry on the {\J}, {\H}, and {\K}
images using Jon Holtzman's very heavily modified version of DAOPHOT
(Stetson 1987). One of the modifications allowed us to fit stars in
the three bands simultaneously, fixing the relative positions of stars
within groups but allowing the offsets between frames to vary. We
carefully added stars to each group by hand and did not allow the
software to delete any. We restricted ourselves to the region of
common overlap between our dithered exposures, to avoid problems
associated with PSF variations (McCaughrean 1993).

Second, we scaled and subtracted the {\Br} line image from each of
the broad band images before performing photometry. The scaling was
determined subjectively. This procedure works well, but it is not
perfect. Clearly, the ratio of broad band nebular emission to {\Br}
emission is not constant, presumably because of the presense of a
component such as scattered light or dust emission that is not
proportional to the emission measure and because of variable
extinction. The effect of this procedure can be seen in
Figure~\ref{fig-jhk} which shows images of the
nebula in {\J}, {\H}, and {\K} both before and after suppression of
the nebular emission.

Our photometry depended to some degree on the scaling factors applied
to the {\Br} line image. We determined this additional uncertainty by
performing photometry on images for which the scaling factors had been
changed by $\pm 10\%$. Since these changes produced images that were
quite noticeably under- or over- subtracted, we assumed that they
correspond to 3$\sigma$ errors. For the ionizing star these errors
amounted to a 0.01--0.02~mag contribution to the 1$\sigma$ error but
for the faintest stars they amounted to as much as a 0.3~mag
contribution to the 1$\sigma$ error.

Our first night's data were taken under better seeing conditions than
the second's, but the second night's data were photometric.
Accordingly, we derived relative magnitudes from the first night's
data and fixed the zero point using aperture photometry of relatively
isolated stars in the second night's data.

Table~\ref{tab-stars} gives astrometry and photometry for all stars in
a 1 arcmin box centered on the ionizing star that have photometric
errors of less than 0.5 in $K$ and 1.0 in both the {\JH} and {\HK}
colors. The errors shown are 1$\sigma$. Table~\ref{tab-stars} is
ordered by increasing $\K$ magnitude and includes the offsets ($\Delta
x$ and $\Delta y$ in arcsec) from the ionizing star. The ionizing star
is \#3 and its close neighbor to the WSW is \#4.

We measure $\K = 10.36 \pm 0.04$, $\JH = 2.40 \pm 0.05$, and $\HK =
1.59 \pm 0.05$ the ionizing star. Lumsden \& Hoare (1996) report $K_n
= 11.2$, $\JH = 2.4$, and $H-K_n = 1.6$ with ``typical errors of
0.3--0.5 mag''. Thus, their colors are in good agreement with ours but
their magnitudes are about 0.8 mag fainter. We assessed the
reliability of our broad band photometry in the following manner. Our
{\Br} flux is in good agreement with other measurements, suggesting
that our narrow band calibration is essentially correct. We measured
2.17~{\micron} fluxes for the isolated stars \#1, \#2, and \#6 using
simple aperture photometry. These fluxes agreed with our {\K}
magnitudes to within 0.02~mag. This suggests that the our {\K}
photometry of isolated stars is essentially correct. We do not believe
we could have made a 0.8~mag error in the relative magnitudes of star
\#3 to the others; for star \#3 to be as faint as $K = 11.2$ would
require it to be almost as faint as star \#5 (located 14 arcsec WSW of
the ionizing star in Figure~\ref{fig-jhk}) but it is clearly much
brighter. Although we cannot offer an explanation for the disagreement
between our magnitudes and those of Lumsden \& Hoare, these
considerations give us confidence that ours are more reliable.

Since the stars were selected by hand, completeness is not well
defined. Furthermore, the completeness varies considerably between
those regions far from the nebula and bright stars, where stars as
faint as $J \approx 17$, $H \approx 16$, and $K \approx 15$ have
$1\sigma$ errors of 0.1 and are easy to detect, to those regions close
to the nebula and bright stars where similarly bright stars might
easily have been missed.

Figure~\ref{fig-ccd} shows color-color diagrams of stars from Table~1.
Only stars with $1\sigma$ errors of less that 0.15 in $K$ and $0.3$ in
both colors are shown. Figure~\ref{fig-ccd}a shows stars within a 30
arcsec square centered on the ionizing star and Figure~\ref{fig-ccd}b
shows stars outside this square. In Figure~\ref{fig-ccd}a, the
ionizing star is shown with a solid square. The error bars show
1$\sigma$ errors. The dotted lines are the loci of unreddened main
sequence and giant stars from Bessell \& Brett (1988) and Koornneef
(1983). The dashed lines are reddening vectors corresponding to
$A_\lambda \propto \lambda^{-1.6}$ (lower), $\lambda^{-1.8}$ (middle),
and $\lambda^{-2.0}$ (upper) extinction laws (Martin \& Whittet 1990).
We have assumed mean wavelengths of 1.25~{\micron}, 1.65~{\micron},
and 2.20~{\micron} for the {\J}, {\H}, and {\K} filters. The length of
each vector corresponds to $A_{\Br} = 2.20$. The quadrilateral
delimits the expected colors for a hot star in the UC~HII region with
$A_{\Br} = 2.20 \pm 0.25$ or $A_K = 2.14 \pm 0.25$. (The scaling
between $A_{\Br}$ and $A_K$ does not depend significantly on the
particular extinction law.) It can be seen that most of the stars have
colors consistent with moderately reddened normal stars.

\subsection{The Ionizing Star}

Van~Buren (1993), Megeath (1993), Fey et~al.\ (1995), and Lumsden \&
Hoare (1996) have suggested that star \#3 is the principal ionizing
source of the UC HII region. Star \#3 has $\K = 10.36 \pm 0.04$, $\JH
= 2.40 \pm 0.05$, and $\HK = 1.59 \pm 0.05$. These colors are in
excellent agreement with those predicted for a hot star reddened by
the measured extinction of the nebula (shown as a
quadrilateral in Figure~\ref{fig-ccd}a). Of the stars that are candidate
members of the young cluster (discussed below), \#3 is the brightest,
has no infrared excess, and is situated in the center of the brightest
arc of nebular emission. This strongly supports the previous
suggestions that it is the primary source of ionization.
There is no evidence for an infrared excess at {\K}. This suggests
that any remaining disk is optically thin at temperatures of about
1300~K and so the star is no longer accreting.

We can estimate the intrinsic apparent {\K} magnitude {\mk} of the
ionizing star by using the extinction of $\AK = 2.14$ with a 3$\sigma$
uncertainty of 0.25 and our measurement of $K = 10.36$ with a
3$\sigma$ uncertainty of 0.12. We find the intrinsic {\mk} is 8.22
with a 3$\sigma$ uncertainty of 0.28. Since the star does not have an
infrared excess, we can estimate the intrinsic apparent {\V} magnitude
{\mv} from {\mk} by assuming an intrinsic {\VK} close to $-0.90$
(Koornneef 1983). We find the intrinsic {\mv} is 7.32 with a 3$\sigma$
uncertainty of 0.28. In Section~\ref{sec-lbol} we determined $3.90 \le
\mbol \le 4.65$ with a $3\sigma$ uncertainty on the limits of 0.3.
To proceed we need to relate {\mv} and {\mbol}. We adopt the
bolometric correction $\BC \equiv \mbol - \mv$ given by Vacca, Garmany,
\& Shull (1996). (We use their equations~3 and 5 and hold the
luminosity class constant at V, as the luminosity class makes very
little difference.) This comes from a fit to a number of non-LTE
atmospheres. 

With our limits on the distance, {\mv}, {\mbol}, and $\mv - \mbol$ we
can construct limits on the {\Lbol} and {\Teff} of the star. The
regions allowed by $1\sigma$, $2\sigma$ and $3\sigma$ errors are shown
in Figure~\ref{fig-model} by thick solid lines. The thin dotted lines
show the loci where $\mv = 7.32$ at different distances. Also shown in
Figure~\ref{fig-model} are evolutionary tracks and isochrones of the
$Z = 2Z_\odot$ models of Meynet et al.\ (1994). The thin solid lines
are tracks for 120, 85, 60, 40, 25, and 20{\Msun} models, the thick
dashed line is the zero-age main-sequence (ZAMS), and thin dashed
lines are the 1, 2, 3, $4 \times 10^6~\rm yr$ isochrones. We only show
the tracks on the first excursion to the red. The dashed-dotted lines
are approximate example stellar birth lines. Stellar birth lines have
not yet been constructed at $Z=2Z_\odot$. Instead, we have applied the
differences in the {\Lbol} and {\Teff} between the Meynet et al.\
(1994) $Z=Z_\odot$ and $Z=2Z_\odot$ ZAMS to the $Z=Z_\odot$ `case 1'
(right) and `case 3' (left) models of Bernasconi \& Maeder (1996).
(The `case 1' and `case 3' models differ in their accretion rate.)
This difference accounts for the change in atmospheric structure, to
first order, but fails to account for the increase of the rate of
evolution with metallicity (Bernasconi 1996, private communication).
These birth lines are, therefore, almost certainly somewhat too blue.

It can be seen from Figure~\ref{fig-model} that the 1, 2, and
3$\sigma$ upper limits on the effective temperature of the ionzing
star are $37\,500$, $40\,000$, and $42\,500\rm~K$. If the ionizing
star is a single star or an unequal-mass binary, our observations
place a firm lower limit on its mass of about 30{\Msun}. If almost all
of the bolometric luminosity in the region arises from the ionizing
star, as we suspect, then this limit is about 60{\Msun}.

The ZAMS lies entirely outside the region allowed by our observations.
Although equal-mass binaries are common among O stars (Garmany, Conti,
\& Massey 1980), a pair of binary ZAMS stars are also inconsistent
with our observations. For consistency with the ZAMS, we require three
ZAMS 40{\Msun} stars at the minimum allowed distance and a conspiracy
of $3\sigma$ errors. We examined the residual image after subtracting
a scaled PSF and found no evidence for elongation. We conservatively
place an upper limit on the separation of the components of such a
system as half the FWHM or about 4000~AU at 5~kpc. This is much
smaller than the size typical for early-type Trapezium systems (Abt
1986). The dynamical time for such a compact system is only about
$10^4~\rm yr$, so it would dissolve over the estimated age of
$10^5~\rm yr$ unless it were formed in a stable, hierarchical state.
We consider this explanation forced and extremely unlikely.

The tracks of evolved stars with ages in excess of $10^6\rm~yr$
coincide with the region allowed by the data, but we reject this
possibility because the age of the UC~HII region is only about
$10^5\rm~yr$. Since the ionizing star appears to be too cool at a
given luminosity to be a ZAMS star, one might think it could be a
pre-main-sequence star evolving onto the ZAMS from the Hayashi track
at roughly constant {\Lbol} but increasing {\Teff}. However, the
modern theory of the formation of massive stars is that
pre-main-sequence stars evolve almost along the ZAMS with both {\Lbol}
and {\Teff} increasing, although they diverge from the ZAMS at the
highest masses because of evolutionary effects (Beech \& Mitalas 1994;
Bernasconi \& Maeder 1996). When they finish accreting, the most
massive stars are somewhat cooler than the ZAMS. This locus is known
as the stellar birth line. Our observations suggest that the stellar
birth line at $Z=2Z_\odot$ must be sufficiently red that the star
could evolve from it into the region allowed by the data in only about
$10^5~\rm yr$. Such a birth line must be cooler than about 40000~K at
least at some point for stars more massive than about 40{\Msun}. Two
example stellar birth lines from Bernasconi \& Maeder (1996), crudely
modified as described above, are shown by dashed-dotted lines in
Figure~\ref{fig-model}. It can be seen that they do indeed diverge
from the ZAMS sufficiently to explain the appearance of a star with an
age of $10^5~\rm yr$ at such low temperatures.

\subsection{A Young Cluster}

Fey et al.\ (1995) suggested that star \#3 is a member of a young
cluster. We are now in a good position to test this suggestion, as we
know the extinction to the nebula and have good colors for the stars.
Unfortunately, we cannot perform a statistical test for an overdensity
of stars close to the UC~HII region because completeness is so complex
and poorly determined in the vicinity of the nebula.

The stars with $\HK > 1$ in Table~\ref{tab-stars} and
Figure~\ref{fig-ccd}a are \#3, \#4, \#13, \#14, and \#24. These stars
are marked in Figure~\ref{fig-members}. Since they are projected close
to the UC HII region and have extinctions that are similar the UC HII
region, these are good candidate members of a cluster that is
physically associated with the UC~HII region. It is worth noting that
with the exception of \#3 and \#14 these stars are not especially
prominent in the images. By a visual inspection of the images alone,
we might well have suggested that the bright stars to the NNE of star
\#3 were part of the cluster, but these appear to be foreground stars
by virtue of their lower extinctions.

The star with the most extreme infrared excess is star \#4, located
only 1.9 arcsec SSW of the ionizing star. This star is difficult to
see without suppressing the nebula emission (see Figure~\ref{fig-jhk})
and has not been previously noted. Comparing it directly to the
ionizing star gives an {\HK} excess of 0.75 with a $3\sigma$ error of
0.19. The presense of this infrared excess star and the absence of
similar stars away from the nebula is further evidence both that the
cluster is real and that it is young.

We estimated the spectral types of these stars from their {\H}
magnitudes as the {\K} magnitudes of the infrared excess stars are
presumably severely contaminated with emission from circumstellar
dust. We scale the extinction to the nebula by a $\lambda^{-1.8}$ law
and find $A_H \approx 3.6$. At a distance of 5~kpc the ionizing star
\#3 has $\MH \approx -5.1$ and the other stars \#4, \#13, \#14, and
\#24 are much fainter with $\MH \approx -3.6$, $-3.0$, $-2.7$, and
$-1.8$ with uncertainties of at least half a magnitude. Cluster stars
other than the ionizing star have luminosities appropriate for early
to mid B stars (Schmidt-Kaler 1982; Koornneef 1983). The situation in
the cluster appears somewhat similar to that in M17 (Hanson \& Conti
1995) in which naked O stars appear with enshrouded B stars, although
possibly here the division between naked and enshrouded stars is not
so sharp.

No stars are visible in the vicinity of the dense core located about 5
arcsec W of the ionizing star. This is consistent with the estimate of
$A_K \approx 400$ through the core (Cesaroni et~al.\ 1994).

\section{Summary and Discussion}

We have demonstrated that the stellar and nebular properties of UC~HII
regions can be studied in the near infrared using moderate integration
times on moderate aperture telescopes. These observations give direct,
quantitative information on the color and luminosity of the ionizing
star. Additionally, they give important information on the extinction
to the nebula and the pattern of dust in and around the nebula.

Our observations of {\g} confirm that the extended radio continuum
emission seen by Fey et~al.\ (1995) is indeed associated with the bright
arc of emission seen by Wood \& Churchwell (1989a). This extended
emission does not appear to be naturally explained by the bow shock
model. Fey et~al.\ (1995) suggest a stationary model in which the arc
and the extended emission are explained by the density
gradient in the molecular cloud. The extinction to the UC~HII
region derived from radio continuum and {\Br} imaging amounts to $\AK
\approx 2.25$ or $\AV \approx 25$ and is higher in the region of the
arc. The most likely explanation for the additional extinction is that
we are seeing a dense molecular sheath around the arc or that there is
significant internal extinction. The required magnitude of the
internal extinction requires roughly twice as much column density in
low density or neutral gas as there is in dense ionized gas in the
vicinity of the arc. This might naturally occur in models in which
UC~HII regions ablate dense inclusions (Dyson 1994; Lizano \& Cant\'o
1995; Dyson, Williams, \& Redman 1995; Redman, Williams, \& Dyson
1996; Williams, Dyson, \& Redman 1996; Lizano et~al.\ 1996) or in
which the arc arises from a photoevaporation front being driven into
an inhomogeneous medium.

We confirm the identity of the ionizing star responsible for ionizing
the UC~HII region. It is located at the center of the arc of emission,
as predicted by both the bow shock and stationary models for the
morphology of the UC~HII region. Direct light from the ionizing star
accounts for only about 1/6 of the total light in each of the {\J},
{\H}, and {\K} bands. The rest is presumably the result of thermal
emission from dust, scattering from dust, and line and continuum
emission from ionized gas.

The ionizing star does not possess a significant excess at {\K},
indicating that it is no longer accreting. This also suggests that the
photoevaporating disk model of Hollenbach et al.\ (1994) may not be
applicable to {\g} in its current state, although detailed
predictions of the spectrum of such a disk are required to confirm
this.

We confirm the suggestion of Fey et~al.\ (1995) that the ionizing star
is closely associated with a number of stars with luminosities
appropriate for B stars. One of these stars has a strong infrared
excess, confirming that the cluster is young. The lack of other
luminous sources suggests that the ionizing star dominates the
bolometric luminosity of the cluster and so is likely to have a mass
of 60{\Msun} or more.

Our measurement of the intrinsic {\mk} of the ionizing star, along
with existing limits on the bolometric luminosity and distance, allow
us to place limits on where the ionizing star can appear in the HR
diagram. We find that the ionizing star is too cool to be within
$10^6\rm~yr$ of the ZAMS, in apparent contradiction with the age of
about $10^5\rm~yr$ estimated for the the UC~HII region. This suggests that
the stellar birth line (Beech \& Mitalas 1994; Bernasconi \& Maeder
1996) for {\g} must be fairly cool: cooler than about 40000~K for
some mass in excess of 40{\Msun}. This places an important limit on
how blue the stellar birth line can be; a red limit already exists, as
the stellar birth line must extend to at least about 90{\Msun}
(Bernasconi \& Maeder 1996). The implications of such a red birth line
for the accretion history cannot be judged at the moment as
theoretical models have not yet been constructed at the twice solar
metallicity appropriate for {\g}. We place a firm lower limit on the
mass of the ionizing star at 30{\Msun} but favor a mass in excess of
about 60{\Msun}.

Our discovery that the ionizing star in {\g} is significantly cooler
than the ZAMS is a further strike against the method of determining
the spectral type of the ionizing star in an UC~HII region from an
estimate of its ionizing continuum flux or spectral hardness. We can
see this by considering the solar metallicity models of Schaerer \& de
Koter (1996). These models combine a stellar atmosphere code and a
stellar evolution code. Their evolved 85{\Msun} and 60{\Msun} models
with $\log \Teff \approx 4.6$ (models E3 and D3) have similar Lyman
continuum fluxes to their ZAMS 40{\Msun} and 25{\Msun} models (models
C1 and B1). Although these specific models have solar metallicity, we
expect that similar discrepancies will arise regardless of
metallicity. Attempts to use the bolometric luminosity to determine
the mass of the ionizing stars in UC~HII regions are in better shape,
as massive stars evolve initially at almost constant {\Lbol}, although
source confusion remains a problem.

\acknowledgements

This work has benefited from discussions with and comments by Paulo
Bernasconi, Pepe Franco, Jay Gallagher, Margaret Hanson, Lynne
Hillenbrand, Melvin Hoare, Todd Hunter, Tom Megeath, \& Robin Williams
and from a very thorough report from an anonymous referee. We thank
Jon Holtzman for access to and instruction in the use of his
simultaneous PSF-fitting photometry routines. Our observations were
made using the Perkins Reflector of the Ohio Wesleyan University and
the Ohio State University at the Lowell Observatory. We thank Andrew
Afflerbach, Riccardo Cesaroni, Mark Claussen, Alan Fey, Ralph Gaume,
and Stan Kurtz for allowing us to use their published and unpublished
data. AMW and PH are grateful for partial support from NATO grant
No.~960776. ALC is grateful for an NSF REU internship administered by
Kathy Degioia-Eastwood and Northern Arizona University.

\pagebreak
\pagestyle{empty}
\setcounter{topnumber}{20}
\setcounter{totalnumber}{20}
\clearpage
\section*{Figure Captions}

\figcaption[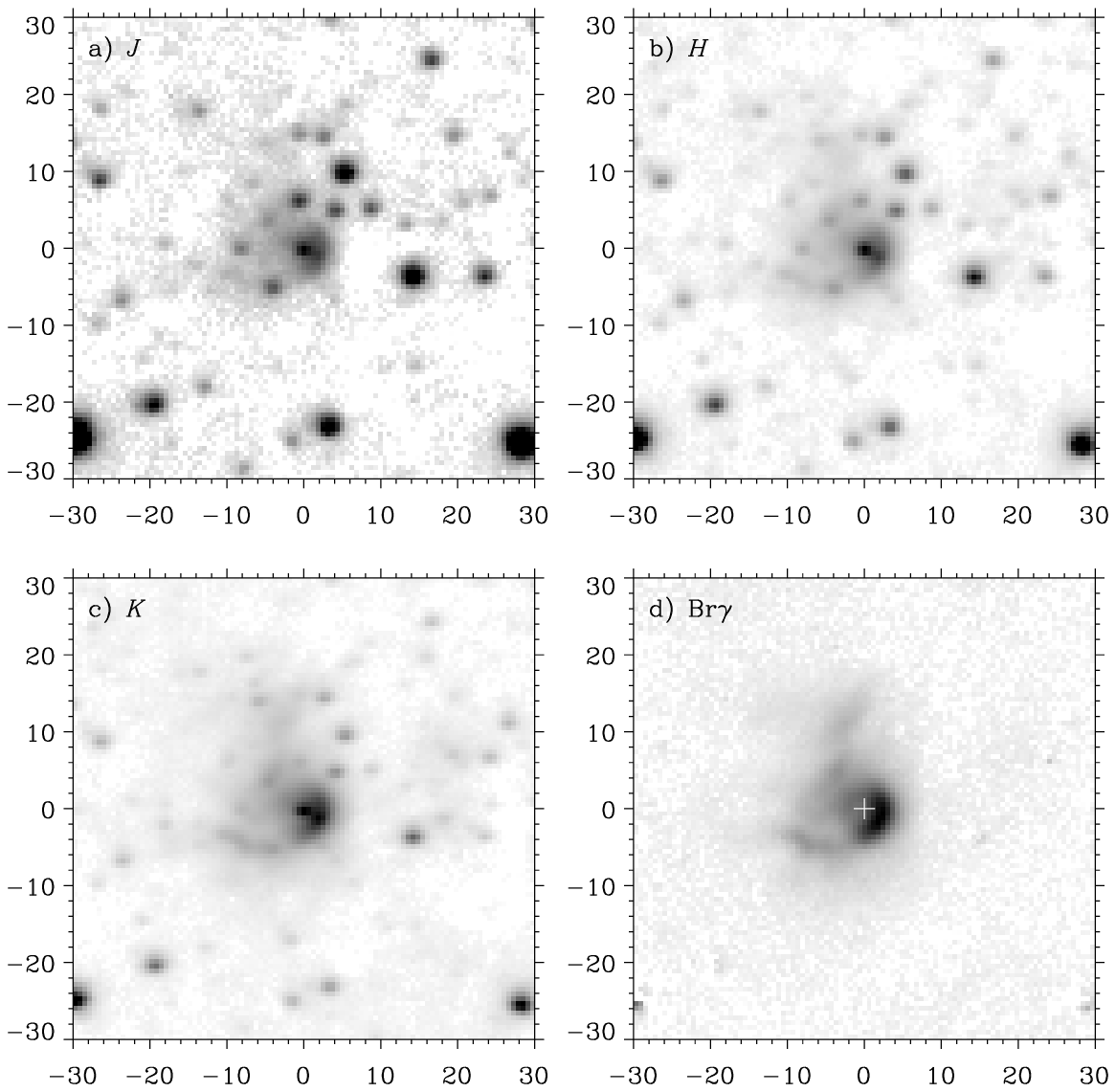]{\label{fig-ir}A $60 \times 60$ arcsec
region around {\g}. The axes are marked in arcsec north and west of
the ionizing star. Darker shades indicate brighter emission. (a) {\J}
band. (b) {\H} band. (c) {\K} band. (d) {\Br} with a white cross
marking the location of the ionizing star.}

\figcaption[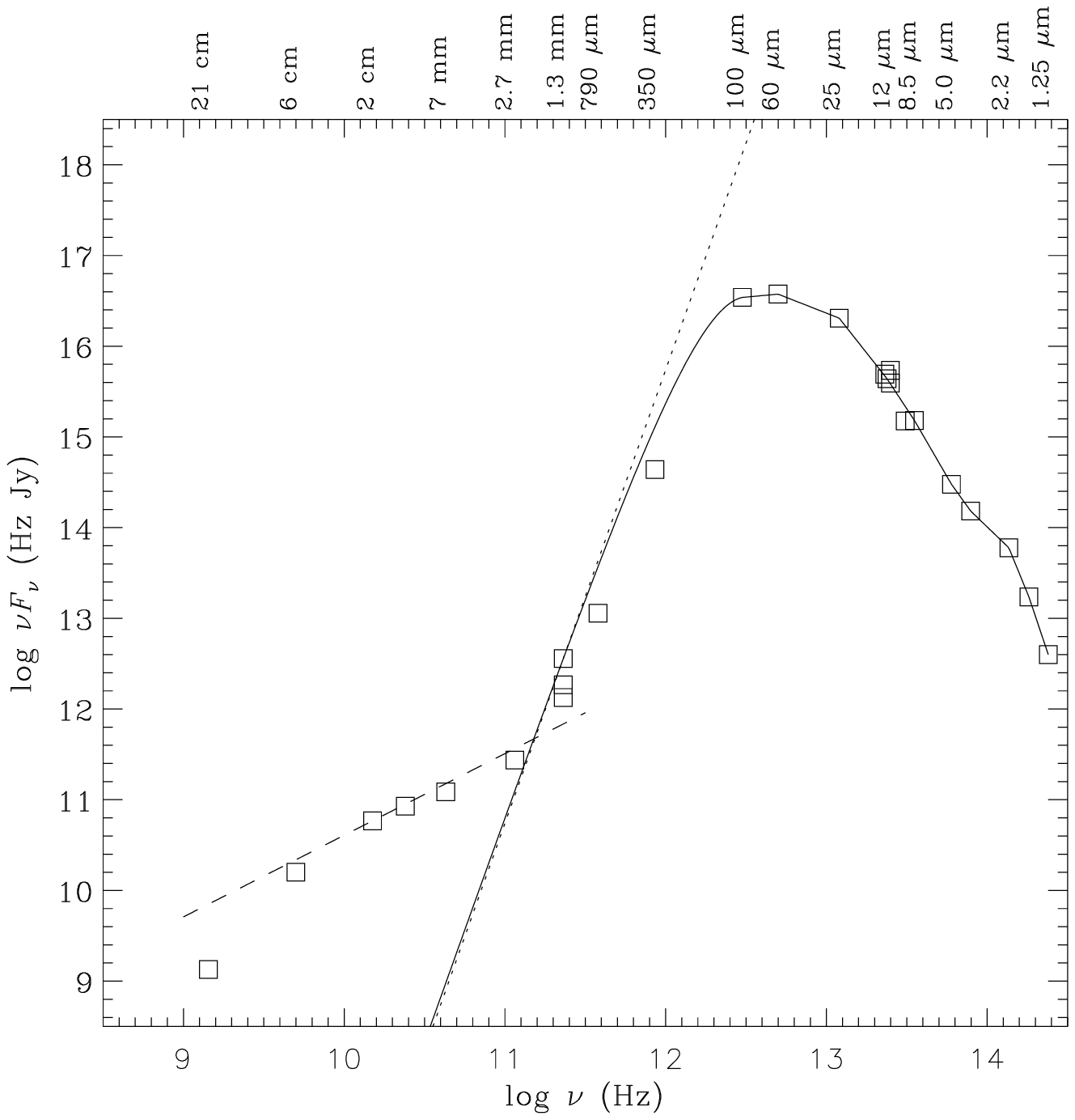]{\label{fig-sed}The spectral energy
distribution for {\g}. The points are taken from
Table~\protect\ref{tab-sed}. The dashed line is a $\nu^{0.1}$ extrapolation from
the 2~cm measurement and shows that optically thin free-free
emission dominates to about 2~mm. The dotted line is a $\nu^4$
extrapolation from the 1300~{\micron} measurement and shows that
the sub-mm region is dominated by optically thin thermal dust
emission. The solid line below 100~{\micron} is a fit
to $\nu^2B_\nu(T)$ which gives $T = 26\rm~K$. The solid line above
100~{\micron} is a linear interpolation in $(\log F_\nu, \log \nu)$.
}

\figcaption[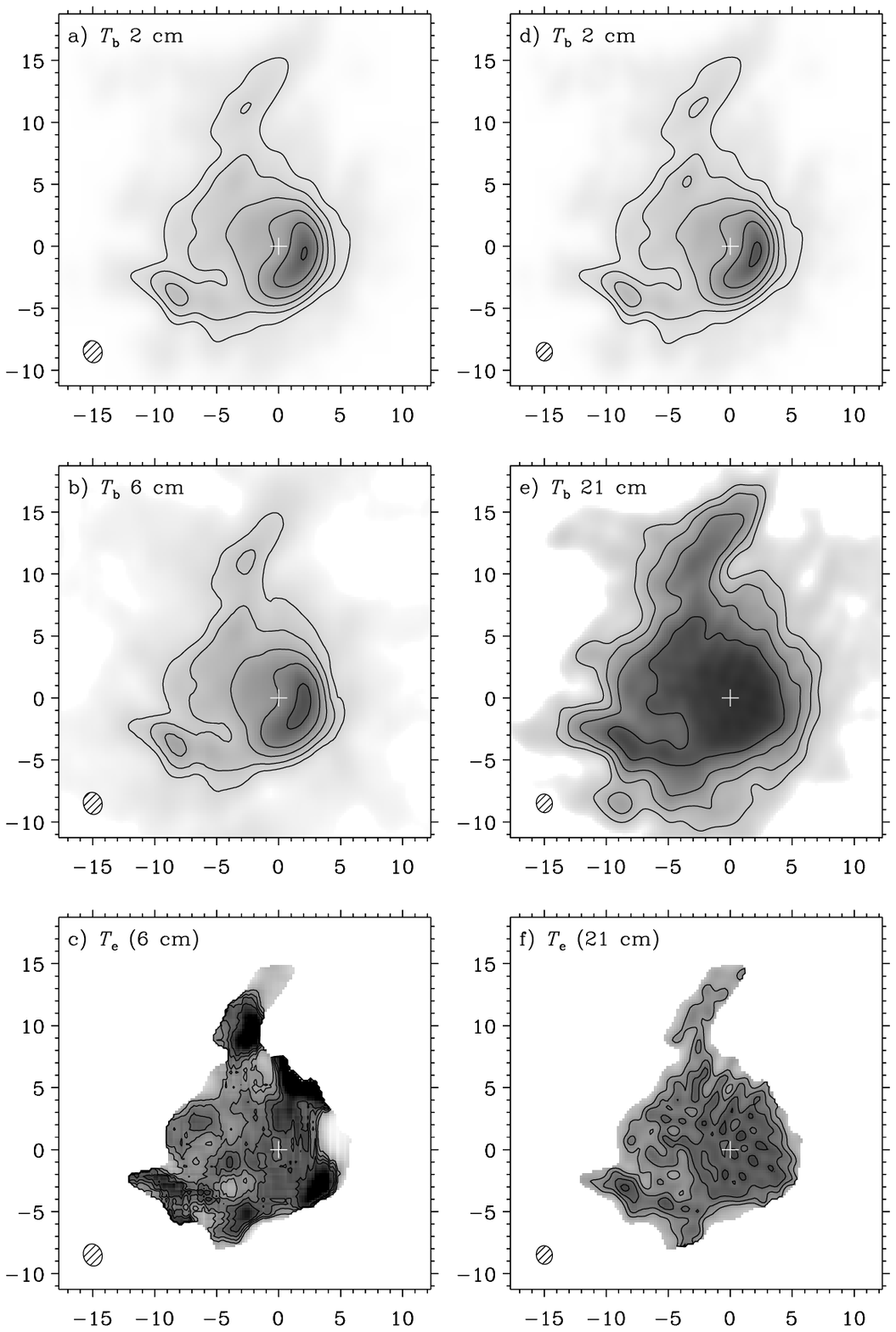]{\label{fig-te} A $30 \times 30$ arcsec region
around {\g}. In each panel, the white cross marks the location of
the ionizing star, the axes are marked in arcsec north and west of the
star, and the beam size is marked in the lower left. Darker shades
indicate higher temperatures. (a) The brightness temperature {\Tb} at
2~cm smoothed to the resolution of the 6~cm image. The contours are at
25, 50, 100, 200, 400, and 800~K. (b) The brightness temperature {\Tb}
at 6~cm. The contours are at 250, 500, 1000, 2000, and 4000~K. (c) The
electron temperature {\Te} derived from the 2~cm and 6~cm images. The
contours are at 4000, 5000, 6000, and 7000~K. (d) The brightness
temperature {\Tb} at 2~cm smoothed to the resolution of the 21~cm
image. The contours are at 25, 50, 100, 200, 400, and 800~K. (e) The
brightness temperature {\Tb} at 21~cm. The contours are at 500, 1000,
2000, and 4000~K. (c) The electron temperature {\Te} derived from the
2~cm and 21~cm images. The contours are at 4000, 5000, 6000, and
7000~K.}

\figcaption[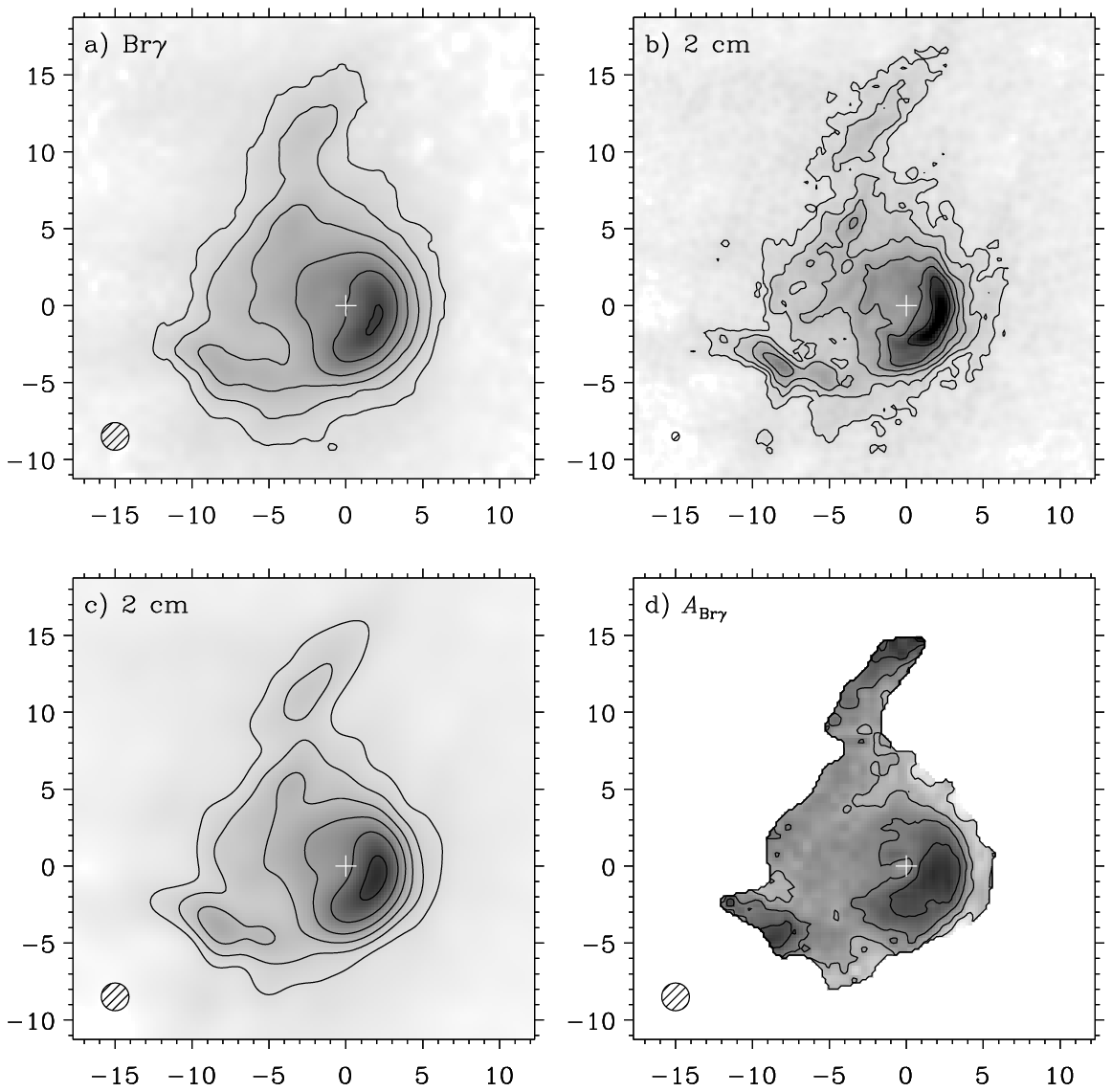]{\label{fig-a}A $30 \times 30$ arcsec region
around {\g}. In each panel, the white cross marks the location of
the ionizing star, the axes are marked in arcsec north and west of the
star, and the beam size is marked in the lower left. Darker shades
indicate higher values. (a) The observed flux in {\Br}. The contours
are spaced by factors of two. (b) The observed flux in 2~cm radio
continuum at full resolution. The contours are spaced by factors of
two. (c) The observed flux in 2~cm radio continuum smoothed to match
the resolution of the {\Br} image. The contours are spaced by factors
of two. (d) The apparent extinction at {\Br}. The contours are at
1.8, 2.0, 2.2, 2.4, and 2.6. }

\figcaption[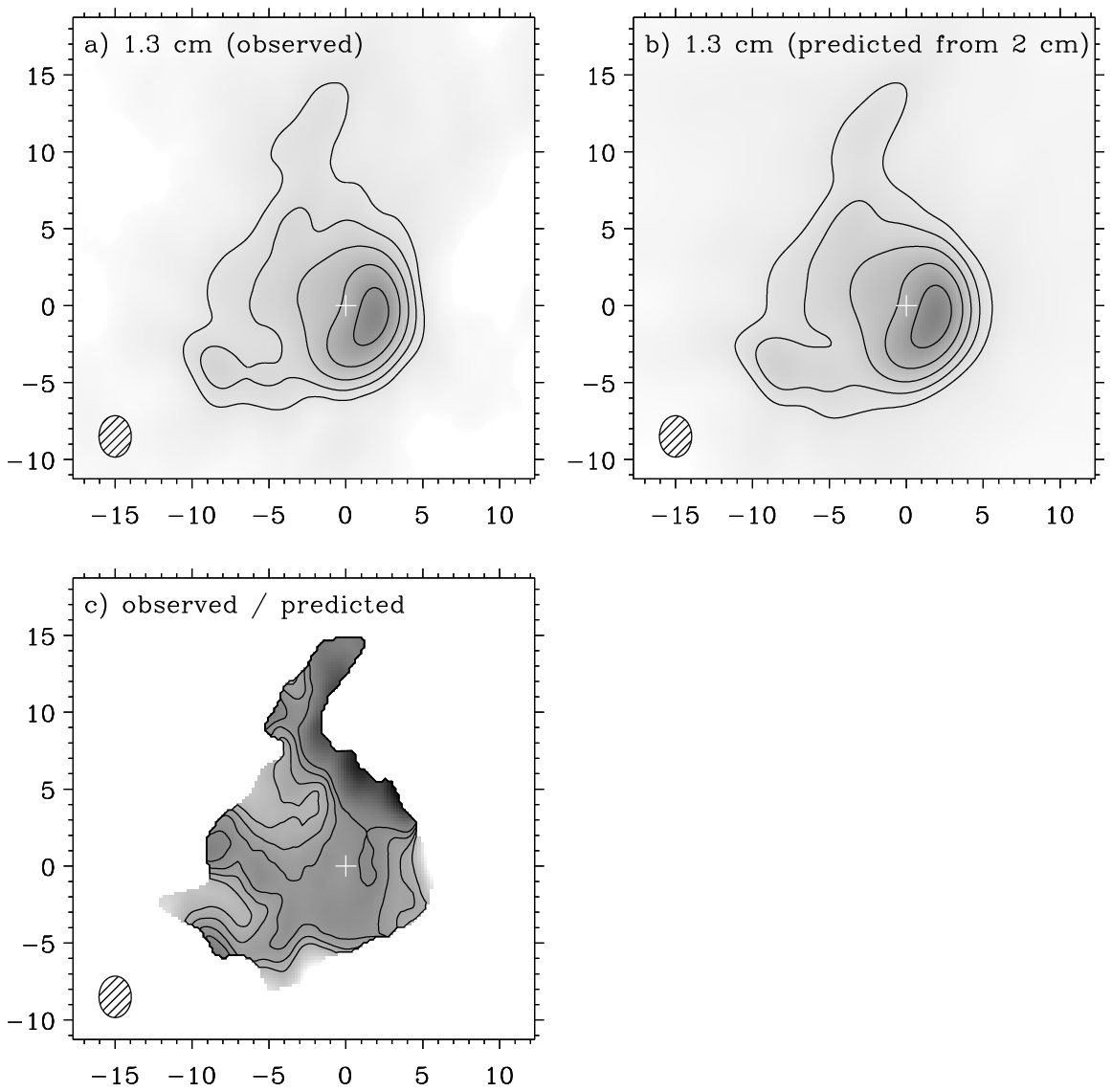]{\label{fig-comp}A $30 \times 30$ arcsec
region around {\g}. In each panel, the white cross marks the
location of the ionizing star, the axes are marked in arcsec north and
west of the star, and the beam size is marked in the lower left.
Darker shades indicate higher values. (a) The observed flux at 1.3~cm.
The contours are spaced by factors of two. (b) The predicted flux at
1.3~cm from the 2~cm image. The contours are spaced by factors of two
and have the same values as in (a). (f) The ratio of the observed flux
to the predicted flux at 1.3~cm. The contours are at 0.8, 0.85, and 0.9}

\figcaption[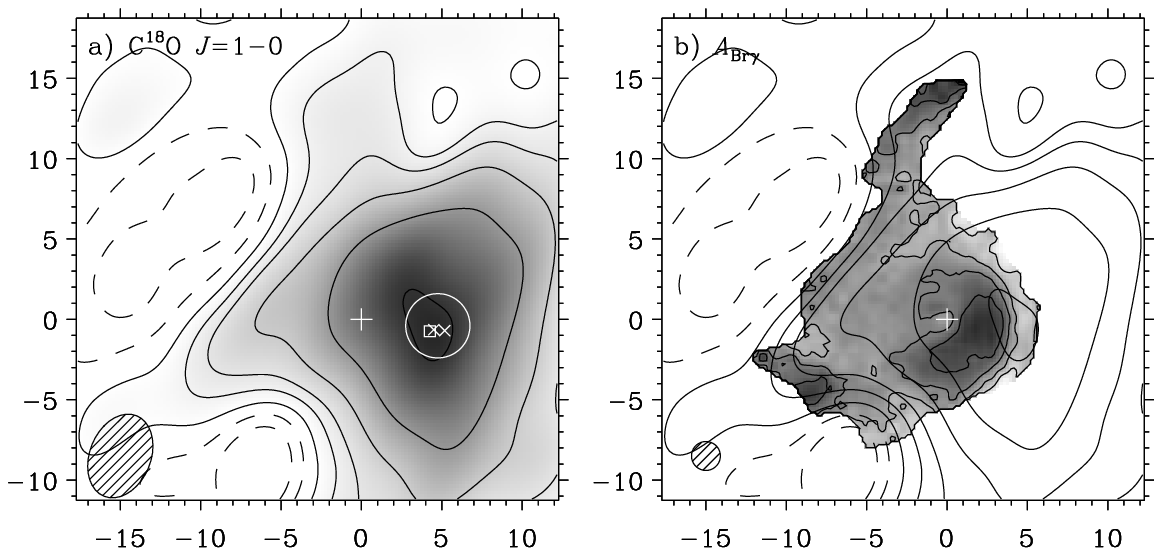]{\label{fig-co}A $30 \times 30$ arcsec region
around {\g}. In each panel, the white cross marks the location of
the ionizing star, the axes are marked in arcsec north and west of the
star, and the beam size is marked in the lower left. Darker shades
indicate higher values. (a) The total line flux in C$^{18}$O. The
contours are spaced by factors of two. Negative contours are dashed.
The white circle marks the FWHM 4 arcsec ammonia clump seen by
Cesaroni et al.\ (1994). The white crosses ($\times$) mark the locations
of the water masers seen by Hofner \& Churchwell (1993). The white
square ($\Box$) marks the locations of the formaldehyde maser seen by
Pratap, Menten, \& Snyder (1994). (b) The apparent extinction at
{\Br}. The contours are at 1.8, 2.0, 2.2, 2.4, and 2.6. The C$^{18}$O
contours from (a) are superposed.}

\figcaption[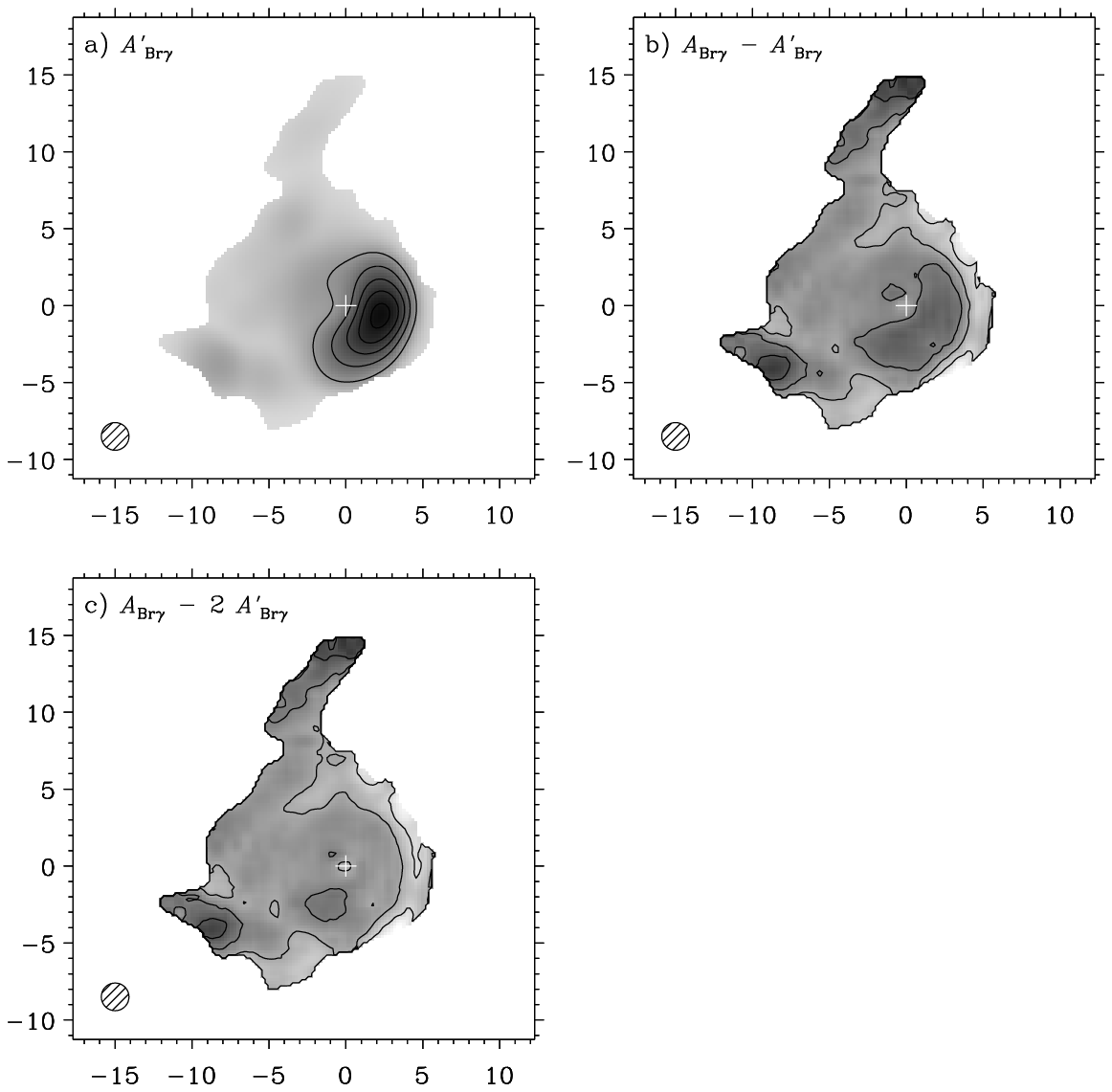]{\label{fig-aint}A $30 \times 30$ arcsec
region around {\g}. In each panel, the white cross marks the
location of the ionizing star, the axes are marked in arcsec north and
west of the star, and the beam size is marked in the lower left.
Darker shades indicate higher values. (a) The model internal
extinction at {\Br}. The contours are at 0.05, 0.10, 0.15, 0.20, and
0.25. (b) The apparent external extinction after removing the model
internal extinction. The contours are at 1.8, 2.0, 2.2, 2.4, and 2.6.
(c) The apparent external extinction after removing twice times the
model internal extinction. The contours are at 1.8, 2.0, 2.2, 2.4, and
2.6. }

\figcaption[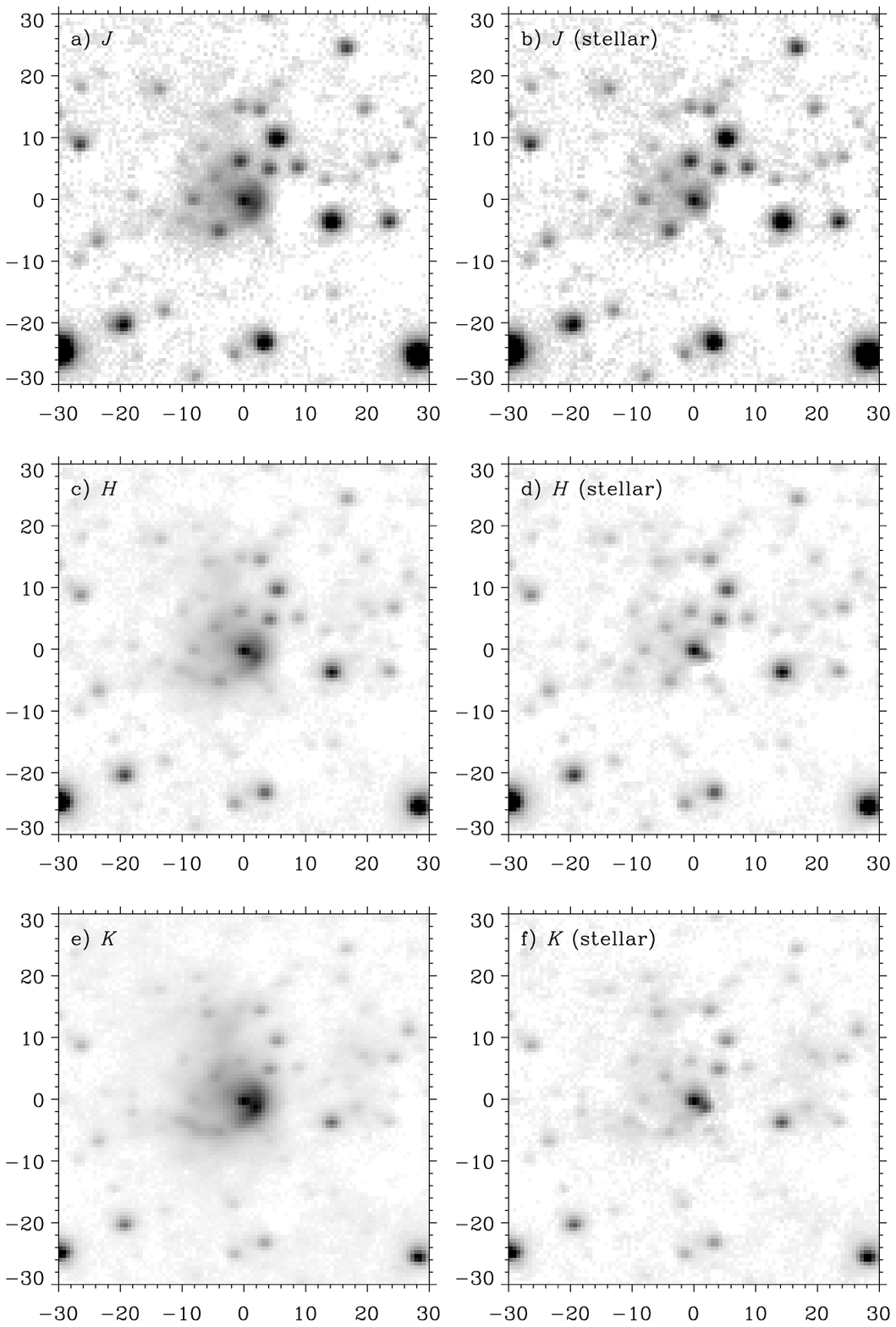]{\label{fig-jhk}A $60 \times 60$ arcsec
region around {\g}. Darker shades indicate brighter emission. (a)
{\J} band. (b) {\J} band with nebular emission suppressed. (c) {\H}
band. (d) {\H} band with nebular emission suppressed. (e) {\K} band.
(f) {\K} band with nebular emission suppressed.}

\figcaption[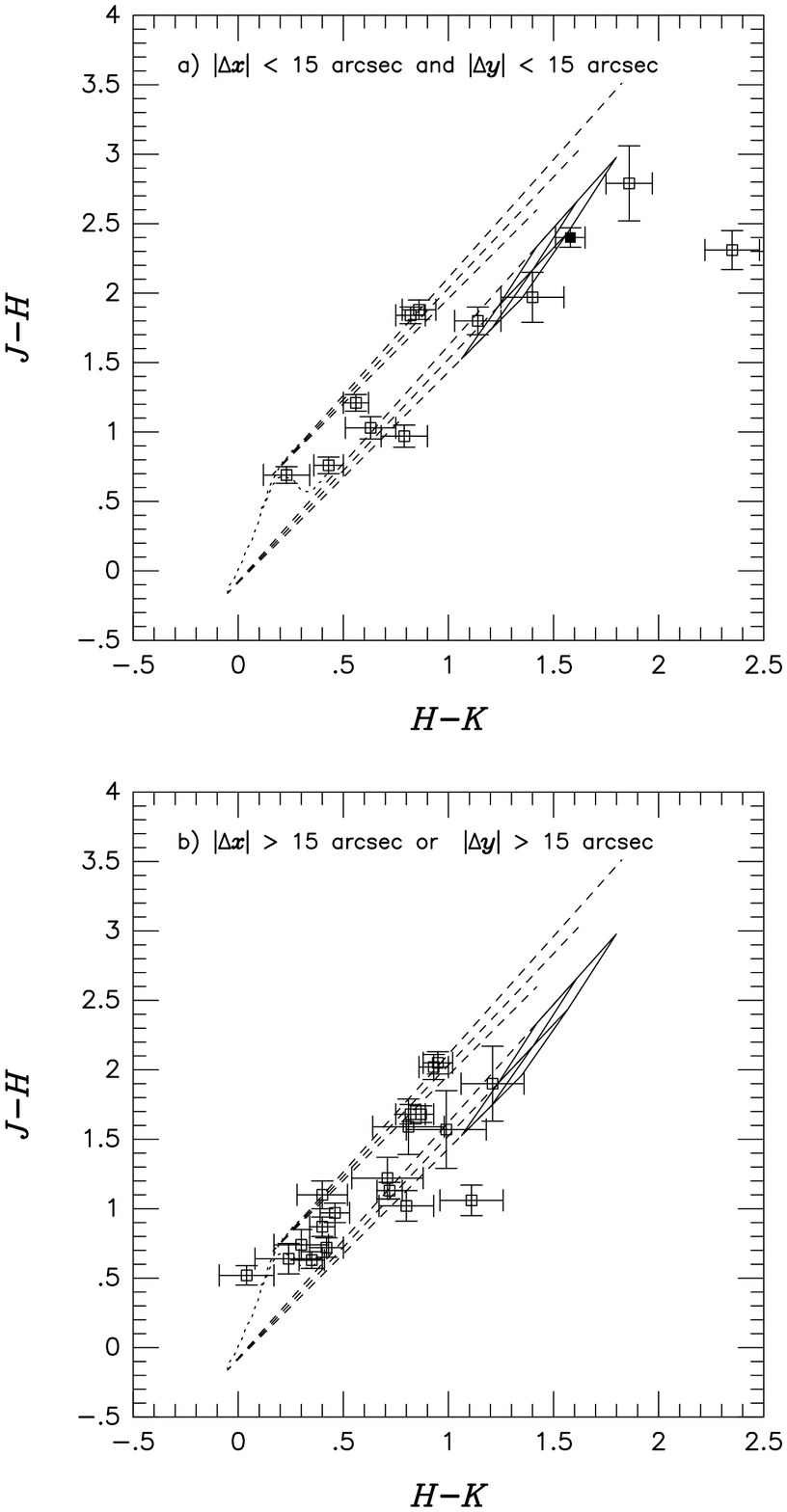]{\label{fig-ccd}(a) Color-color diagram for stars
within a 30 arcsec square centered on the ionizing star. (b)
Color-color diagrams for stars outside that box. The error bars are
1$\sigma$. Only stars with $1\sigma$ errors of less that 0.15 in $K$
and $0.3$ in both colors are shown. The ionizing star is marked with a
solid symbol. The dotted lines are the loci of main-sequence and giant
stars. The dashed lines are reddening vectors of $\lambda^{-1.6}$,
$\lambda^{-1.8}$, and $\lambda^{-2.0}$ reddening laws and correspond
to $A_{\Br} = 2.20$. The quadrilateral shows the expected colors of a
hot star with $A_{\Br} = 2.20 \pm 0.25$ under these reddening laws. }

\figcaption[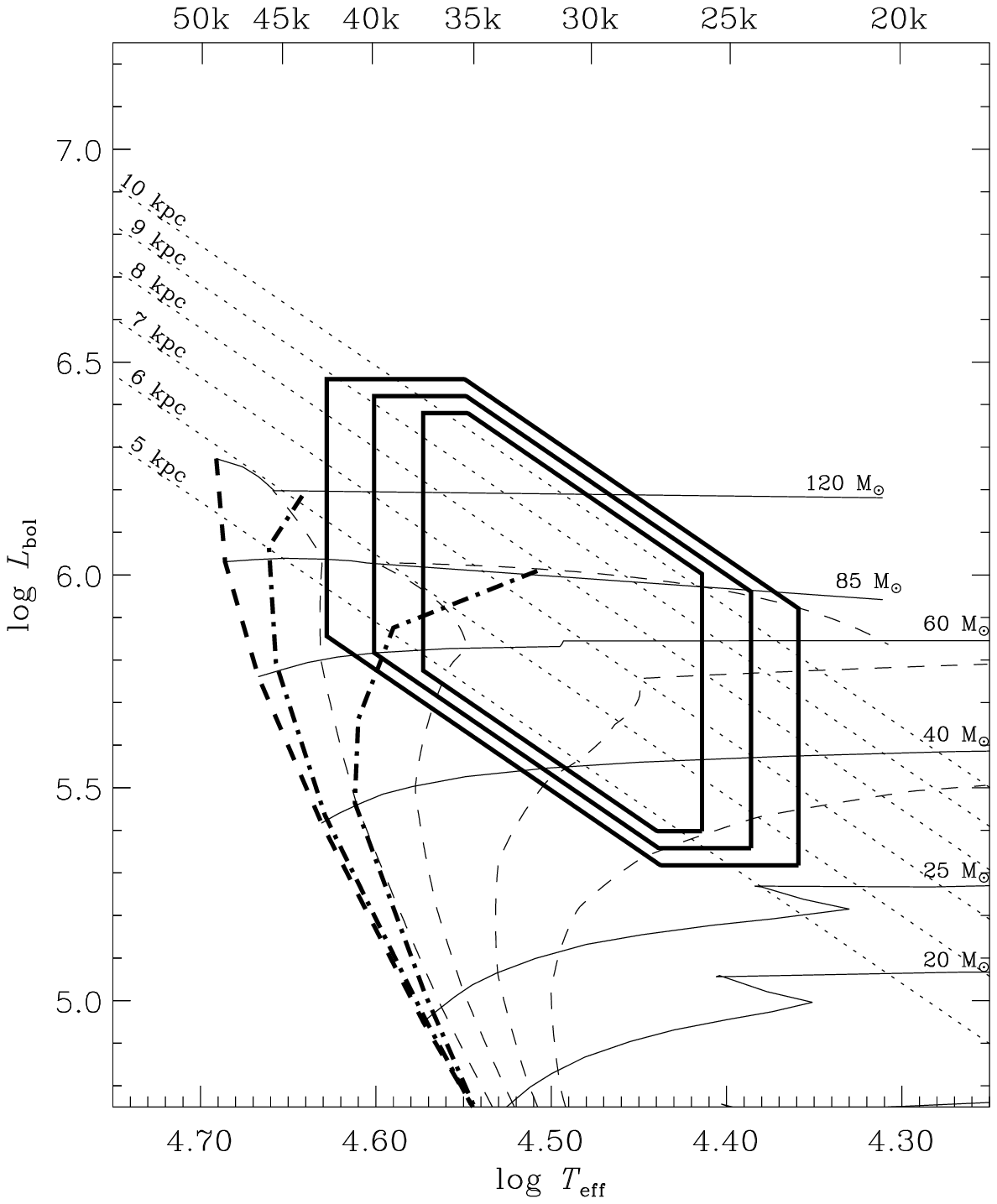]{\label{fig-model}A theoretical HR
diagram. The region allowed by $1\sigma$, $2\sigma$, and $3\sigma$
uncertainties on the measured properties of the ionizing star are
shown by thick solid lines. The thin dotted lines show the loci where
$\mv = 7.32$ at different distances. Also shown are tracks and isochrones of
the $Z = 2Z_\odot$ models of Meynet et al.\ (1994). The solid lines
are tracks for 120, 85, 60, 40, 25, and 20{\Msun} models, the dashed
lines are the zero-age main-sequence (ZAMS), and 1, 2, 3, $4 \times
10^6~\rm yr$ isochrones. The dashed-dotted lines are approximate
stellar birth lines. As described in the text, these birthlines are
almost certainly somewhat too blue.}

\figcaption[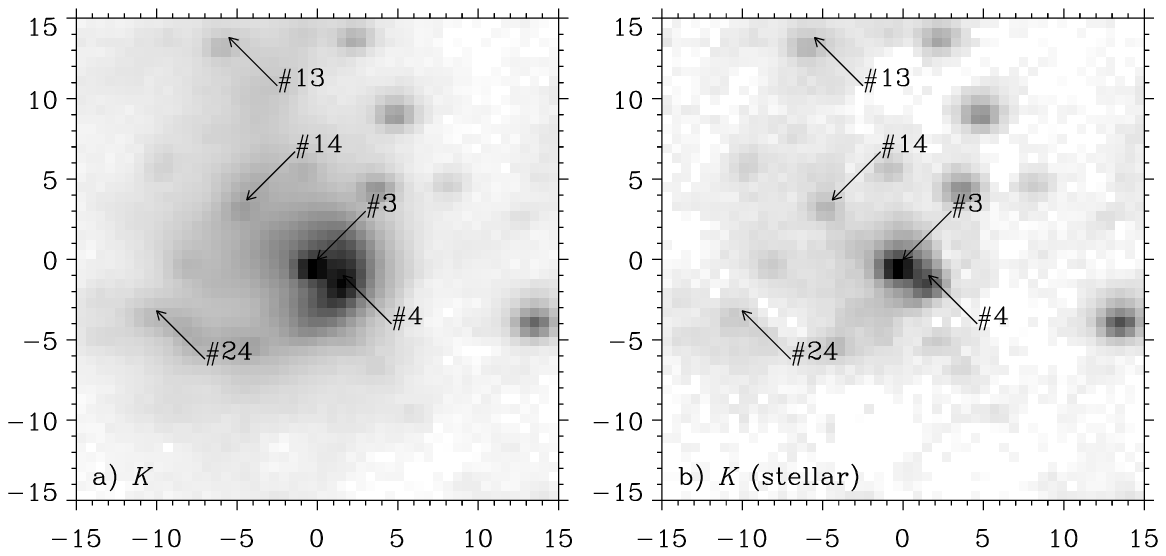]{\label{fig-members} (a) A $30 \times 30$ arcsec
region around {\g} showing the {\K} band emission. Darker shades indicate brighter emission.
Candidate members of a cluster associated with the UC~HII region are
marked. (b) As (a) but with nebular emission suppressed.}

\begin{figure}
\leftline{Fig.~\ref{fig-ir}}
\plotone{./f1.eps}
\end{figure}

\begin{figure}
\leftline{Fig.~\ref{fig-sed}}
\plotone{./f2.eps}
\end{figure}

\begin{figure}
\leftline{Fig.~\ref{fig-te}}
\plotone{./f3.eps}
\end{figure}

\clearpage
\begin{figure}
\leftline{Fig.~\ref{fig-a}}
\plotone{./f4.eps}
\end{figure}

\clearpage
\begin{figure}
\leftline{Fig.~\ref{fig-comp}}
\plotone{./f5.eps}
\end{figure}

\clearpage
\begin{figure}
\leftline{Fig.~\ref{fig-co}}
\plotone{./f6.eps}
\end{figure}

\clearpage
\begin{figure}
\leftline{Fig.~\ref{fig-aint}}
\plotone{./f7.eps}
\end{figure}

\clearpage
\begin{figure}
\leftline{Fig.~\ref{fig-jhk}}
\plotone{./f8.eps}
\end{figure}

\clearpage
\begin{figure}
\leftline{Fig.~\ref{fig-ccd}}
\plotone{./f9.eps}
\end{figure}

\clearpage
\begin{figure}
\leftline{Fig.~\ref{fig-model}}
\plotone{./f10.eps}
\end{figure}

\clearpage
\begin{figure}
\leftline{Fig.~\ref{fig-members}}
\plotone{./f11.eps}
\end{figure}

\pagebreak
\pagestyle{empty}
\setcounter{topnumber}{20}
\setcounter{totalnumber}{20}
\clearpage

\section*{Table Captions}

\begin{table}[h]
\caption{\label{tab-radio-data}Radio Data}
\end{table}

\begin{table}[h]
\caption{\label{tab-sed}Spectral Energy Didstribution}
\end{table}

\begin{table}[h]
\caption{\label{tab-stars}Stellar Photometry}
\end{table}

\setcounter{table}{0}
\pagebreak

\begin{deluxetable}{lllcc}
\tablewidth{\textwidth}
\tablecaption{Radio Data}
\tablehead{
\colhead{Authors}&
\colhead{Frequency}&
\colhead{Telescope}&
\colhead{Beam}&
\colhead{LAS}\nl
&&&
\colhead{arcsec}&
\colhead{arcsec}
}
\startdata
Fey et al.\ (1995)		&14.9~GHz/2~cm		&VLA B+C+D	&$0.6 \times 0.5$	&90\nl
Cesaroni et al.\ (1994)	&24.1~GHz/1.3~cm	&VLA C+D	&$2.7 \times 2.1$	&60\nl
Afflerbach et al.\ (1994)	&4.87~GHz/6~cm	&VLA B		&$1.8 \times 1.5$	&40\nl
Claussen \& Hofner (1996)	&1.43~GHz/21~cm	&VLA A		&$1.3 \times 1.5$	&40\nl
Hofner et al.\ (1996)		&110~GHz/2.7~mm	&OVRO		&$3.8 \times 5.4$	&20\nl
\enddata
\end{deluxetable}

\clearpage

\begin{deluxetable}{lll}
\tablewidth{\textwidth}
\tablecaption{Spectral Energy Distribution}
\tablehead{
\colhead{Wavelength}&
\colhead{$F_\nu$}&
\colhead{Reference}
\nl
&
\colhead{Jy}
}
\startdata
21~cm			&\phantom{0}0.94					&Claussen \& Hofner (1996)	\\
6~cm			&\phantom{0}3.18					&Afflerbach et al.\ (1994)	\\
2~cm			&\phantom{0}3.9						&Fey et al.\ (1995)			\\
1.3~cm			&\phantom{0}3.53					&Cesaroni et al.\ (1994)	\\
7~mm			&\phantom{0}2.85					&Wood et al.\ (1988)		\\
2.7~mm			&\phantom{0}2.37					&Hofner et al.\ (1997)		\\
1.3~mm			&15.6\tablenotemark{a}				&Mooney et al.\ (1995)		\\
1.3~mm			&\phantom{0}8.0						&Chini et al.\ (1986)		\\
1.3~mm			&\phantom{0}5.81\tablenotemark{b}	&Mooney et al.\ (1995)		\\
790~\micron		&\phantom{0}$5.1 \times 10^2$		&Hunter (1997)				\\
350~\micron		&\phantom{0}$3.0 \times 10^1$		&Hunter (1997)				\\
100~\micron		&\phantom{0}$1.15 \times 10^4$		&IRAS PSC					\\
60~\micron		&\phantom{0}$7.50 \times 10^3$		&IRAS PSC					\\
25~\micron		&\phantom{0}$1.70 \times 10^3$		&IRAS PSC					\\
13.0~\micron	&\phantom{0}$2.13 \times 10^2$		&Ball et al.\ (1996)		\\
12~\micron		&\phantom{0}$2.17 \times 10^2$		&IRAS PSC					\\
12.6~\micron	&\phantom{0}$1.84 \times 10^2$		&Ball et al.\ (1996)		\\
12.0~\micron	&\phantom{0}$1.57 \times 10^2$		&Ball et al.\ (1996)		\\
%ms: \tablebreak
9.7~\micron		&\phantom{0}$4.84 \times 10^1$		&Ball et al.\ (1996)		\\
8.5~\micron		&\phantom{0}$4.30 \times 10^1$		&Ball et al.\ (1996)		\\
5.0~\micron		&\phantom{0}$5.0$					&Herter et al.\ (1981)		\\
3.8~\micron		&\phantom{0}$1.9$					&Herter at al.\ (1981)		\\
2.2~\micron		&\phantom{0}$4.40 \times 10^{-1}$	&This work					\\
1.65~\micron	&\phantom{0}$9.48 \times 10^{-2}$	&This work					\\
1.25~\micron	&\phantom{0}$1.66 \times 10^{-2}$	&This work					\\
\enddata
\tablenotetext{a}{Integrated flux}
\tablenotetext{b}{12 arcsec FWHM point source}
\end{deluxetable}

\clearpage

\pagebreak
\begin{deluxetable}{rrrrrr}
\tablewidth{\textwidth}
\tablecaption{Stellar Photometry}
\tablehead{
\colhead{Star}&
\colhead{$\Delta x$}&
\colhead{$\Delta y$}&
\colhead{\JH}&
\colhead{\HK}&
\colhead{\K}\nl
&
\colhead{arcsec}&
\colhead{arcsec}&
}
\startdata
 1 & $-28.9$ & $-23.9$ & $+0.87 \pm 0.05$ & $+0.41 \pm 0.05$ & $9.61 \pm 0.04$ \\
 2 & $+27.4$ & $-24.6$ & $+0.98 \pm 0.06$ & $+0.47 \pm 0.06$ & $10.12 \pm 0.04$ \\
 3 & $ +0.0$ & $ +0.0$ & $+2.40 \pm 0.05$ & $+1.59 \pm 0.05$ & $10.36 \pm 0.04$ \\
 4 & $ +1.6$ & $ -1.0$ & $+2.28 \pm 0.11$ & $+2.34 \pm 0.07$ & $11.14 \pm 0.05$ \\
 5 & $+13.8$ & $ -3.4$ & $+1.22 \pm 0.05$ & $+0.57 \pm 0.05$ & $11.41 \pm 0.03$ \\
 6 & $-18.8$ & $-19.5$ & $+1.68 \pm 0.05$ & $+0.89 \pm 0.04$ & $11.55 \pm 0.03$ \\
 7 & $ +3.9$ & $ +4.8$ & $+1.84 \pm 0.05$ & $+0.83 \pm 0.05$ & $12.34 \pm 0.04$ \\
 8 & $ +5.1$ & $ +9.5$ & $+0.76 \pm 0.05$ & $+0.44 \pm 0.05$ & $12.35 \pm 0.04$ \\
 9 & $ +3.2$ & $-22.3$ & $+0.63 \pm 0.04$ & $+0.37 \pm 0.05$ & $12.46 \pm 0.03$ \\
10 & $ +2.5$ & $+14.3$ & $+1.89 \pm 0.06$ & $+0.86 \pm 0.06$ & $12.92 \pm 0.05$ \\
11 & $-25.6$ & $ +8.7$ & $+1.13 \pm 0.05$ & $+0.74 \pm 0.05$ & $12.93 \pm 0.03$ \\
12 & $ -1.4$ & $-24.0$ & $+2.02 \pm 0.08$ & $+0.95 \pm 0.06$ & $13.10 \pm 0.04$ \\
13 & $ -5.5$ & $+13.8$ & $+2.79 \pm 0.26$ & $+1.86 \pm 0.09$ & $13.20 \pm 0.05$ \\
14 & $ -4.4$ & $ +3.7$ & $+1.80 \pm 0.09$ & $+1.15 \pm 0.08$ & $13.23 \pm 0.06$ \\
15 & $ -3.8$ & $ -4.8$ & $+0.98 \pm 0.06$ & $+0.79 \pm 0.08$ & $13.41 \pm 0.06$ \\
16 & $+23.4$ & $ +6.7$ & $+2.04 \pm 0.07$ & $+0.97 \pm 0.06$ & $13.42 \pm 0.05$ \\
17 & $-22.9$ & $ -6.3$ & $+1.69 \pm 0.06$ & $+0.83 \pm 0.06$ & $13.47 \pm 0.05$ \\
18 & $ -2.4$ & $ +1.1$ & $+1.34 \pm 0.15$ & $+1.66 \pm 0.14$ & $13.56 \pm 0.11$ \\
19 & $+16.2$ & $+23.9$ & $+0.72 \pm 0.06$ & $+0.44 \pm 0.07$ & $13.61 \pm 0.05$ \\
20 & $ -2.3$ & $ -2.9$ & $+1.92 \pm 0.22$ & $+1.80 \pm 0.17$ & $13.68 \pm 0.13$ \\
\tablebreak
21 & $+17.6$ & $ -1.1$ & $+2.28 \pm 0.38$ & $+2.18 \pm 0.21$ & $13.68 \pm 0.09$ \\
22 & $+18.8$ & $ +7.2$ & $+2.84 \pm 0.41$ & $+2.07 \pm 0.11$ & $13.75 \pm 0.08$ \\
23 & $ -0.6$ & $ +6.1$ & $+0.70 \pm 0.05$ & $+0.23 \pm 0.09$ & $13.79 \pm 0.08$ \\
24 & $-10.0$ & $ -3.2$ & $+1.97 \pm 0.17$ & $+1.39 \pm 0.13$ & $13.93 \pm 0.10$ \\
25 & $+22.7$ & $ -3.3$ & $+0.52 \pm 0.06$ & $+0.08 \pm 0.12$ & $14.01 \pm 0.11$ \\
26 & $ -7.8$ & $ +0.1$ & $+1.04 \pm 0.07$ & $+0.62 \pm 0.11$ & $14.03 \pm 0.10$ \\
27 & $-29.3$ & $+13.4$ & $+1.38 \pm 0.08$ & $+0.83 \pm 0.07$ & $14.06 \pm 0.05$ \\
28 & $-17.6$ & $ -1.9$ & $+1.14 \pm 0.40$ & $+2.90 \pm 0.27$ & $14.25 \pm 0.08$ \\
29 & $+15.3$ & $+19.2$ & $+1.58 \pm 0.33$ & $+1.84 \pm 0.20$ & $14.36 \pm 0.08$ \\
30 & $+29.5$ & $+28.5$ & $+0.95 \pm 0.09$ & $+0.48 \pm 0.09$ & $14.41 \pm 0.07$ \\
31 & $ +8.4$ & $ +5.1$ & $+0.79 \pm 0.06$ & $-0.03 \pm 0.17$ & $14.48 \pm 0.16$ \\
32 & $ -9.4$ & $ +6.2$ & $+2.25 \pm 0.22$ & $+0.95 \pm 0.16$ & $14.49 \pm 0.14$ \\
33 & $ +3.5$ & $+29.0$ & $+1.01 \pm 0.11$ & $+0.82 \pm 0.12$ & $14.49 \pm 0.10$ \\
34 & $-13.2$ & $+17.5$ & $+1.11 \pm 0.09$ & $+0.42 \pm 0.12$ & $14.53 \pm 0.10$ \\
35 & $+20.0$ & $ +6.0$ & $+1.07 \pm 0.10$ & $+1.13 \pm 0.14$ & $14.56 \pm 0.12$ \\
36 & $-20.3$ & $-13.8$ & $+1.89 \pm 0.26$ & $+1.23 \pm 0.15$ & $14.83 \pm 0.10$ \\
37 & $+29.1$ & $ +1.3$ & $+1.36 \pm 0.11$ & $+0.72 \pm 0.11$ & $14.83 \pm 0.08$ \\
38 & $-26.4$ & $+23.2$ & $+2.36 \pm 0.53$ & $+1.43 \pm 0.18$ & $14.87 \pm 0.11$ \\
39 & $-27.3$ & $+10.1$ & $+1.58 \pm 0.19$ & $+0.85 \pm 0.16$ & $14.91 \pm 0.12$ \\
40 & $ +0.7$ & $-22.0$ & $+1.48 \pm 0.26$ & $+1.06 \pm 0.18$ & $14.94 \pm 0.12$ \\
\tablebreak
41 & $-12.4$ & $-17.3$ & $+0.74 \pm 0.10$ & $+0.32 \pm 0.13$ & $14.98 \pm 0.10$ \\
42 & $-25.9$ & $ -9.3$ & $+1.23 \pm 0.15$ & $+0.73 \pm 0.16$ & $15.01 \pm 0.13$ \\
43 & $+19.1$ & $+14.4$ & $+0.64 \pm 0.10$ & $+0.26 \pm 0.16$ & $15.06 \pm 0.13$ \\
44 & $ -0.6$ & $+14.6$ & $+0.90 \pm 0.10$ & $+0.03 \pm 0.29$ & $15.24 \pm 0.28$ \\
45 & $-20.5$ & $-19.7$ & $+0.70 \pm 0.12$ & $+0.18 \pm 0.24$ & $15.34 \pm 0.22$ \\
46 & $ -7.6$ & $-27.4$ & $+0.62 \pm 0.12$ & $+0.28 \pm 0.19$ & $15.42 \pm 0.17$ \\
47 & $-17.6$ & $ +1.1$ & $+1.15 \pm 0.11$ & $+0.34 \pm 0.26$ & $15.52 \pm 0.25$ \\
48 & $+14.2$ & $-14.5$ & $+1.20 \pm 0.12$ & $+0.35 \pm 0.20$ & $15.53 \pm 0.18$ \\
49 & $ +3.9$ & $ -6.2$ & $+2.10 \pm 0.34$ & $+0.48 \pm 0.43$ & $15.60 \pm 0.41$ \\
50 & $-21.5$ & $-17.9$ & $+1.61 \pm 0.45$ & $+1.22 \pm 0.28$ & $15.61 \pm 0.19$ \\
51 & $-25.6$ & $+17.8$ & $+0.70 \pm 0.11$ & $+0.14 \pm 0.20$ & $15.65 \pm 0.17$ \\
\enddata
\end{deluxetable}

\end{document}